\documentclass[12pt]{article}
\usepackage{amsfonts}

\usepackage{graphicx}
\usepackage{amsmath}
\usepackage{float}

\textwidth=6.5in \hoffset=-.75in \voffset=-1in \numberwithin{equation}{section}
\numberwithin{figure}{section}

\input{tcilatex}

\begin{document}

\begin{titlepage}
\bigskip \begin{flushright}
hep-th/0408058\\
\end{flushright}
\vspace{1cm}
\begin{center}
{\Large \bf {
A Review of the N--bound and the Maximal Mass Conjectures Using
NUT--Charged dS Spacetimes}}\\
\end{center}
\vspace{2cm}
\begin{center}
R. Clarkson{ \footnote{ EMail: r2clarks@sciborg.uwaterloo.ca}}, A.M.
Ghezelbash{ \footnote{ EMail: amasoud@avatar.uwaterloo.ca}}, R. B. Mann{
\footnote{ EMail: mann@avatar.uwaterloo.ca}}\\
Department of Physics, University of Waterloo, \\
Waterloo, Ontario N2L 3G1, Canada\\
\vspace{1cm}
\end{center}

\begin{abstract}
The proposed dS/CFT correspondence remains an intriguing paradigm in the
context of string theory. Recently it has motivated two interesting
conjectures: the entropic $\mathbf{N}$--bound and the maximal mass
conjecture. The former states that there is an upper bound to the entropy in
asymptotically de Sitter spacetimes, given by the entropy of pure de Sitter
space. The latter states that any asymptotically de Sitter spacetime cannot
have a mass larger than the pure de Sitter case without inducing a
cosmological singularity.

Here we review the status of these conjectures and demonstrate their
limitation. We first describe a generalization of gravitational
thermodynamics to asymptotically de Sitter spacetimes, and show how to
compute conserved quantities and gravitational entropy using this formalism.
From this we proceed to a discussion of the $\mathbf{N}$--bound and maximal
mass conjectures. We then illustrate that these conjectures are not
satisfied for certain asymptotically de Sitter spacetimes with NUT charge.
We close with a presentation of explicit examples in various spacetime
dimensionalities.
\end{abstract}
\end{titlepage}\onecolumn      
\bigskip


\section{Introduction}

One of the most fundamental features of a physical system are the conserved
quantities that are associated with the system. Asymptotically flat (aF) or
asymptotically Anti de Sitter (aAdS) spacetimes have generally well
understood conserved quantities, though for the aAdS case the situation is
problematic due to the supertranslation--like ambiguities involved. However
in either case the conserved quantities have been defined relative to an
auxiliary spacetime, embedding the boundary of the spacetime of interest
into a reference spacetime. This is not always possible, however, even for
basic spacetimes such as the Kerr solution \cite{Martinez}. Conformal
invariance further restricts the choice these reference spacetimes \cite{CCM}%
, constraining the applicability of this approach.

The AdS/CFT inspired counter--term method has led to a great deal of
progress in this area, as this method no longer requires the use of a
reference spacetime. Instead, additional surface terms, functionals of the
geometric invariants on the boundary of the spacetime \cite{balakraus}, are
used and provide an alternative approach to calculating the desired
conserved quantities of aAdS spacetimes, an approach not plagued by the
above difficulties.

A logical extension of any conjectured relationship in an aAdS spacetime is
of course its counterpart in the asymptotically de Sitter (adS) spacetimes.
The construction of the conserved charges using AdS/CFT counter--terms,
however, depends on the spatial infinity present in any aAdS spacetime, a
feature not shared by adS spacetimes. These spacetimes also have no global
timelike Killing vector, as the norm of the Killing vector changes sign as
it crosses the horizon. Inside the horizon, the Killing vector is timelike,
and this has been used to calculate the conserved charges and
actions/entropies for pure and asymptotically dS spacetimes inside the
cosmological horizon \cite{GibbonsHawking1}. Outside of this cosmological
horizon, however, with the change to a spacelike Killing vector, the
physical meaning of the conserved quantities is unclear; for example, one
could use the conformal Killing vector to calculate the energy if one chose %
\cite{Shiro}.

Recently a method for computing conserved charges (and associated boundary
stress tensors) of adS spacetimes from data at early or late time infinity
was proposed \cite{bala}, in analogy with Brown-York prescription used in
asymptotically AdS spacetimes \cite{balakraus,brown,BCM,ivan}. The result is
a holographic duality similar to the AdS/CFT correspondence, appropriately
called the dS/CFT correspondence. The specific prescription (which has been
employed previously by others but in more restricted contexts \cite%
{Klem,Myung}) used counterterms on spatial boundaries at early and late
times, yielding a finite action for asymptotically dS spacetimes in $3,4,5$
dimensions. By carrying out the procedure, analogous to the AdS cases
already calculated \cite{balakraus,BCM}, the boundary stress tensor on the
spacetime boundary can be calculated, and a conserved charge (spacelike, due
to its association with the Killing vector $\partial /\partial t$, and now
interpreted as the mass \cite{bala}) of the adS can also be calculated.

With this definition in mind, Balasubramanian \textit{et al. }\cite{bala}
were led to what we call the maximal mass conjecture, which states that 
\textit{any asymptotically dS spacetime with mass greater than dS has a
cosmological singularity.} This conjecture is in need of clarification;
however roughly speaking it means that the conserved mass of any physically
reasonable adS spacetime must be negative (i.e. less than the zero value of
pure dS spacetime). The conjecture has been verified for topological dS
solutions and their dilatonic variants \cite{cai}, and for the
Schwarzschild-de Sitter (SdS) black hole \cite{GM}.

It is to be noted that this mass conjecture was based on another conjecture
concerning adS spacetimes, the Bousso $\mathbf{N}$--bound \cite{bousso}. The
Bousso $\mathbf{N}$--bound states that \textit{any asymptotically dS
spacetime will have an entropy no greater than the entropy }$\pi \ell ^{2}$ 
\textit{of pure dS with cosmological constant }$\Lambda =3/\ell ^{2}$\textit{%
\ in }$(3+1)$ \textit{dimensions}.

There is a class of spacetimes --- Taub--NUT--dS spacetimes --- that provide
counterexamples to both of these conjectures \cite{dsnutshort}. Our purpose
here is to review these results in more detail, providing background for
their derivation and indication of how the conjectures are violated. We note
in passing that a class of stable higher dimensional ($d+1=p+q>4$)
spacetimes of the form $dS_{p}\otimes S^{q}$ have entropy greater than dS
spacetime and thus also violate the $\mathbf{N}$--bound \cite{boussomyers}.

In section \ref{sec:conservedcharges}, we will outline and review the
procedure for calculating the conserved mass, angular momentum and entropy
in any asymptotically (A)dS spacetime. In section \ref{sec:counter-term}, we
will present the procedure for deriving boundary counterterms from the
Gauss--Codacci equation for the asymptotically dS spacetime. We show that
these counterterms are sufficient for obtaining a finite action for the
inflationary patches (big bang and big crunch patches) of dS\ spacetime in
any dimensionality. In section \ref{sec:pathintegral}, we review the path
integral formalism in asymptotically dS spacetimes, contrasting it with the
procedure in asymptotically AdS spacetimes. In sections \ref{sec:N-bound}
and \ref{sec:maxmass}, we will review in brief the Bousso $\mathbf{N}$%
--bound on the entropy and the maximal mass conjecture of asymptotically dS
spacetimes. In section \ref{sec:cosmo-spaceimes}, we will consider the
different asymptotically dS spacetimes with rotation and NUT charge and
discuss the regions outside the cosmological horizon. In sections \ref%
{sec:examples} and \ref{sec:examplesC}, we will present our detailed results
for the violations of the Bousso $\mathbf{N}$--bound and maximal mass
conjectures in four dimensional Taub-Bolt-dS spacetimes, using two methods
for computing the path integral and conserved quantities at past/future
infinity. These two methods are equivalent to one another via an analytic
continuation \cite{RCstelea}. Finally, in section \ref{sec:generalcalc}, we
present the general expressions for the conserved mass and the entropy of
the Taub--NUT/Bolt--dS spacetimes in ($d+1$) dimensions and show that
NUT-charged spacetimes of dimensionality $4k$ are qualitatively similar to
the ($3+1$) dimensional case whereas those of dimensionality $4k+2$ are
qualitatively similar to the ($5+1$) dimensional case (for both the R-- and
C--approaches).

\section{Conserved charges and Gravitational Thermodynamics}

\bigskip \label{sec:conservedcharges}

Thermodynamic properties of black holes have been studied extensively for
the past three decades. In these studies, more attention has been paid to
the thermodynamic properties of black holes in asymptotically flat
spacetimes. There are several reasons for extending this framework to
non-asymptotically flat spacetimes, including string-theoretic motivations
connected with asymptotically Anti de Sitter spacetimes and the mounting
empirical evidence that our spacetime has a positive cosmological constant.

In this section, we review the definition of the conserved charges of the
gravitational field within a region of space with boundary $B$, for a
spacetime with negative cosmological constant $\Lambda $. We consider the
spacetime $M$ of dimension $d+1$, which is the product of a spacelike
hypersurface $\Sigma $ and a real line time interval $I$. We denote the
boundary of $\Sigma $ by $B$, and so the boundary of the spacetime $M$
consists of initial and final spacelike hypersurfaces $\Sigma _{i},\Sigma
_{f}$ at $t_{i}$ and $t_{f}$, respectively and a timelike hypersurface $%
\Upsilon =B\otimes I$. The hypersurface $\Upsilon $ joins the hypersurfaces $%
\Sigma _{i}$ and $\Sigma _{f}$. The spacetime metric is $g_{\mu \nu }$ and
we denote the induced metrics on $\Sigma _{i},\Sigma _{f}$ by $h_{ij}$ and
on $\Upsilon $ by $\gamma _{ij}$. The gravitational action is 
\begin{equation}
I=\frac{1}{16\pi G}\int_{\mathcal{M}}d^{d+1}x\sqrt{-g}\left( R-2\Lambda +%
\mathcal{L}_{M}\right) +\frac{1}{8\pi G}\int_{\Sigma _{i}}^{\Sigma
_{f}}d^{d}x\sqrt{h}K-\frac{1}{8\pi G}\int_{\Upsilon }d^{d}x\sqrt{-\gamma }%
\Theta  \label{actiontot}
\end{equation}%
where $\mathcal{L}_{M}$ refers to the matter Lagrangian, which we shall not
consider here. The functions $K$ and $\Theta $ are the traces of the
extrinsic curvatures $K_{ij}$ and $\Theta _{ij}$ for the boundary
hypesurfaces $\Sigma $ and $\Upsilon $, with respect to the future and
outward unit normals, respectively. By using the ADM decomposition of the
boundary metric $\gamma _{ij},$ the timelike boundary hypersurface $\Upsilon 
$ foliates into $d$-dimensional hypersurfaces $B$. In fact, the induced
metric $\sigma _{ab}$ on $B$ is related to $\gamma _{ij}$ by 
\begin{equation}
\sigma _{ab}(dx^{a}+V^{a}dt)(dx^{b}+V^{b}dt)=\gamma
_{ij}dx^{i}dx^{j}+N^{2}dt^{2}  \label{ADM}
\end{equation}%
where $N$ is the lapse function and $V^{a\text{ }}$is the shift vector. The
gravitational stress-energy tensors in the boundaries could be obtained by
variation of the action, under the variation of the boundary metrics. Then
by using the known relation for the variation of the boundary metric $\gamma
_{ij}$ in terms of the variations of the lapse function, shift vector and
induced metric $\sigma _{ab}$, we obtain the following expressions for the
energy surface density $\varepsilon $, momentum surface density $j_{a\text{ }%
}$ and spatial stress $s^{ab}$, which are the normal and tangential
projections of $p^{ij}$ on the boundary $B$: 
\begin{eqnarray}
\varepsilon &=&\frac{2}{N\sqrt{\sigma }}u_{i}p^{ij}u_{j}+\varepsilon _{0} 
\notag \\
j_{a} &=&-\frac{2}{N\sqrt{\sigma }}\sigma _{ai}p^{ij}u_{j}+j_{a0}
\label{densities} \\
s^{ab} &=&\frac{2}{N\sqrt{\sigma }}p^{ab}+s_{0}^{ab}.  \notag
\end{eqnarray}%
In the relations (\ref{densities}), $p^{ij}$ is the energy-momentum tensor
of the boundary $\Upsilon $ (the projection components of the $p^{ij}$ from $%
\Upsilon $ to $B$ is denoted by $p^{ab}$) and $u_{i\text{ }}$is the unit
normal to the hypersurface $B$. The origin of the additive terms $%
\varepsilon _{0}$, $j_{a0}$ and $s_{0}^{ab}$ is from the addition of an
extra term, a functional of the metric on the boundary to the action (\ref%
{actiontot}). Such a term leaves the classical equations of motion
invariant. Variation of this extra term yields additional contributions to
the energy and momentum surface densities and spatial stress. These
quantities could be selected to obtain favorable finite results for the
energy, momentum and stress in the background method or the finite boundary
condition method \cite{BCM}. In the counterterm method, these quantities are
obtained by the variation of the counterterm Lagrangian, which serves to
eliminate divergences in the energy, angular momentum and boundary stress
tensor.

Another equivalent representation of the above equations (\ref{densities})
is available, in terms of the extrinsic curvature of the hypersurface $B$,
the gravitational energy--momentum tensor of the hypersurface $\Sigma $ and
the acceleration of the unit normal to this hypersurface \cite{brown}. The
total quasilocal energy is the integral of the energy surface density over
the $(d-1)$--boundary $B$, 
\begin{equation}
E=\int_{B}d^{d-1}x\sqrt{\sigma }\varepsilon .  \label{en}
\end{equation}%
Moreover, if we have a symmetry on the boundary $\Upsilon $, generated by a
Killing vector $\xi ^{i},$ then the quantity%
\begin{equation}
\mathcal{Q}_{\xi }=\int_{B}d^{d-1}x\sqrt{\sigma }u_{i}p^{ij}\xi _{j}
\label{cc1}
\end{equation}%
is conserved. In the absence of matter, the conserved $\mathcal{Q}_{\xi }$
could be written in terms of energy and momentum surface densities via 
\begin{equation}
\mathcal{Q}_{\xi }=\int_{B}d^{d-1}x\sqrt{\sigma }(\varepsilon
u_{i}+j_{i})\xi ^{i}  \label{cc}
\end{equation}%
where conservation means that $\mathcal{Q}_{\xi }$ is independent of the
particular boundary $B$ that is chosen for its actual calculation. As an
example, consider a case where a timelike Killing vector $\xi ^{i}$, with
unit length, exists. Moreover, we assume that this Killing vector is surface
forming. Then it is the unit normal to a particular foliation of the
boundary $\Upsilon $ and on each slice of this foliation $\xi ^{i}=u^{i}$
and $j_{i}\xi ^{i}=0.$ In this case, the total energy of the slice (\ref{en}%
) is just the negative of $\mathcal{Q}_{\xi }$. For more general slices that
are not orthogonal to the Killing vector, the total energy is different from 
$-\mathcal{Q}_{\xi }$, which defines the conserved mass of the gravitational
system.

\section{Counter-term approach and dS/CFT Correspondence}

\label{sec:counter-term}

In $d+1$ dimensional spacetimes with positive cosmological constant, the
equations of motion can be derived from the action 
\begin{equation}
I=I_{B}+I_{\partial B}  \label{action}
\end{equation}%
where 
\begin{eqnarray}
I_{B} &=&\frac{1}{16\pi G}\int_{\mathcal{M}}d^{d+1}x\sqrt{-g}\left(
R-2\Lambda \right)  \label{actbulk} \\
I_{\partial B} &=&-\frac{1}{8\pi G}\int_{\mathcal{\partial M}^{-}}^{\mathcal{%
\ \ \partial M}^{+}}d^{d}x\sqrt{\gamma ^{\pm }}\Theta ^{\pm }.
\label{actbound}
\end{eqnarray}

The first term (\ref{actbulk}) is the bulk action over the $d+1$ dimensional
Manifold $\mathcal{M}$ with Newtonian constant $G$ and the second term (\ref%
{actbound}) is the Gibbons-Hawking surface term, necessary to ensure a well
defined Euler-Lagrange variation. $\mathcal{\partial M}^{\pm }$ are spatial
Euclidean boundaries at early and late times and $\int_{\mathcal{\partial M}%
^{-}}^{\mathcal{\partial M}^{+}}d^{d}x$ indicates an integral over the late
time boundary minus an integral over the early time boundary. The quantities 
$g_{\mu \nu },\gamma_{\mu \nu }^{\pm }$ and $\Theta^{\pm }$ are the bulk
spacetime metric, induced boundary metrics and the trace of extrinsic
curvatures of the boundaries respectively. We shall usually suppress the ``$%
\pm $'' notation when it is obvious. However, as is well known the action (%
\ref{action}) is not finite when evaluated on a solution of the equations of
motion. The reason is the infinite volume of the spacetime at early and late
times.

The procedure, prior to the use of counter--terms, for dealing with such
divergences in asymptotically flat/AdS cases (where they are large-distance
effects) was to include a reference action term \cite{brown,BCM}, which
corresponded to the action of embedding the boundary hypersurface $\mathcal{%
\partial M}$ (whose unit normal is spacelike) into some other manifold. The
physical interpretation is that one has a collection of observers located on
the closed manifold $\mathcal{\ \partial M}$, and that the physical
quantities they measure (energy, angular momentum, etc.) are those contained
within this closed manifold relative to those of some reference spacetime
(regarded as the ground state) in which $\mathcal{\partial M}$ is embedded %
\cite{nonortho}. For example in an asymptotically Anti de Sitter spacetime,
it would be natural to take pure AdS as the ground state reference manifold.

However this procedure suffers from several drawbacks: the reference
spacetime in general cannot be uniquely chosen \cite{CCM} nor can an
arbitrary boundary $\mathcal{\partial M}$ always be embedded in a given
reference spacetime. Employing approximate embeddings can lead to ambiguity,
confusion and incompleteness; examples of this include the Kerr \cite%
{Martinez}, Taub-NUT and Taub-Bolt spacetimes \cite{TNUT}. Extensions to
asymptotically dS spacetimes are even more problematic, since the embedding
surface will typically be time-dependent.

An alternative approach for asymptotically AdS spacetimes was suggested a
few years ago that has enjoyed a greater measure of success \cite%
{balakraus,hensken,korea,sergey}. It involves adding to the action terms
that depend only on curvature invariants that are functionals of the
intrinsic boundary geometry. Such terms cannot alter the equations of motion
and, since they are divergent, offer the possibility of removing divergences
that arise in the action (\ref{action}) provided the coefficients of the
allowed curvature invariants are correctly chosen. No embedding spacetime is
required, and computations of the action and conserved charges yield
unambiguous finite values that are intrinsic to the spacetime. This has been
explicitly verified for the full range of type-D asymptotically AdS
spacetimes, including Schwarzschild-AdS, Kerr-AdS, Taub-NUT-AdS,
Taub-Bolt-AdS, and Taub-Bolt-Kerr-AdS \cite{EJM,misner,nutkerr}.

The boundary counterterm action is universal, and a straightforward
algorithm has been constructed for generating it \cite{KLS}. The procedure
involves rewriting the Einstein equations in Gauss--Codacci form, and then
solving them in terms of the extrinsic curvature functional $\Pi
_{ab}=\Theta _{ab}-\Theta \gamma _{ab}$ of the boundary $\mathcal{\partial M}
$ and its normal derivatives to obtain the divergent parts. It succeeds
because all divergent parts can be expressed in terms of intrinsic boundary
data, and do not depend on normal derivatives \cite{Feffgraham}. By writing
the divergent part $\tilde{\Pi}_{ab}$ as a power series in the inverse
cosmological constant the entire divergent structure can be covariantly
isolated for any given boundary dimension $d$; by varying the boundary
metric under a Weyl transformation, it is straightforward to show that the
trace $\tilde{\Pi}$ is proportional to the divergent boundary counterterm
Lagrangian.

Explicit calculations have demonstrated that finite values for the action
and conserved charges can be unambiguously computed up to $d=8$ for the
class of Kerr--AdS metrics \cite{saurya}. The removal of divergences is
completely analogous to that which takes place in quantum field theory by
adding counterterms which are finite polynomials in the fields. The AdS/CFT
correspondence conjecture asserts that these procedures are one and the
same. Corroborative evidence for this is given by calculations which
illustrate that the trace anomalies and Casimir energies obtained from the
two different descriptions are in agreement for known cases \cite%
{hensken,korea,adscftcalcs}.

Generalizations of the counterterm action to asymptotically flat spacetimes
have also been proposed \cite{misner,Lau}. They are quite robust, and allow
for a full calculation of quasilocal conserved quantities in the Kerr
solution \cite{dehghani} that go well beyond the slow-rotating limit that
approximate embedding techniques require \cite{Martinez}. Although they can
be inferred for general $d$ by considering spacetimes of special symmetry,
they cannot be algorithmically generated, and are in general dependent upon
the boundary topology \cite{KLS}.

Turning next to the asymptotically de Sitter case, we must add to the action
(\ref{action}) some counterterms to cancel its divergences 
\begin{equation}
I_{ct}=-\frac{1}{8\pi G}\int_{\mathcal{\partial M}^{+}}d^{d}x\sqrt{\gamma } 
\mathcal{L}_{ct}-\frac{1}{8\pi G}\int_{\mathcal{\partial M}^{-}}d^{d}x\sqrt{
\gamma }\mathcal{L}_{ct}  \label{counter}
\end{equation}
so that 
\begin{equation}
I=I_{B}+I_{\partial B}+I_{ct}  \label{totaction}
\end{equation}
is now the total action.

For the special cases $d=2,3,4,$ the counterterm Lagrangian 
\begin{equation}
\mathcal{L}_{ct}=-\frac{d-1}{\ell }+\frac{\ell \Theta \left( d-3\right) }{%
2(d-2)}\hat{R}  \label{count345}
\end{equation}%
was proposed, where $\hat{R}$ is the intrinsic curvature of the boundary
surfaces and the step function $\Theta (x)$ is equal to zero unless $x>0%
\mathbf{,}$ where it equals unity. The action (\ref{count345}) was shown to
cancel divergences in de Sitter spacetime \cite{bala} 
\begin{equation}
ds^{2}=-d\tau ^{2}+\ell ^{2}\exp \left( 2\tau /\ell \right) d\vec{x}\cdot d%
\vec{x}  \label{dsinf}
\end{equation}%
and Nariai spacetime 
\begin{equation}
ds^{2}=-\left( d\frac{\tau ^{2}}{\ell ^{2}}-1\right) ^{-1}d\tau ^{2}+\left( d%
\frac{\tau ^{2}}{\ell ^{2}}-1\right) dt^{2}+\ell ^{2}\left( 1-\frac{2}{d}%
\right) d\Omega _{d-1}^{2}  \label{Narai}
\end{equation}%
where the metric $d\vec{x}\cdot d\vec{x}$ is a flat $d$-dimensional metric
that covers an inflationary patch of de Sitter spacetime and $d\Omega
_{d-1}^{2}$ is the metric of a unit $(d-1)$-sphere. Here 
\begin{equation}
\Lambda =\frac{d\left( d-1\right) }{2\ell ^{2}}  \label{cosmo}
\end{equation}%
is the positive cosmological constant.

Calculations of the total action are performed by cutting off de Sitter
space at a finite time, and then letting the surface approach future
infinity. In odd dimensions, there is an additional divergence that is
logarithmic in the conformal time (that is $\ell \exp \left( -\tau /\ell
\right) $) and cannot be cancelled without including an explicit cutoff
dependence in the counterterm action, thereby leading to a conformal anomaly
similar to what has been observed in the context of AdS spacetimes \cite%
{Confanom,Confanom2}. Furthermore, if we use the global covering coordinates
of the dS spacetimes in which equal time hypersurfaces are $d$--spheres
(instead of inflationary coordinates in which equal time hypersurfaces are
flat), there is (only for odd dimensional de Sitter spacetimes) a linear
divergence in the action that cannot be removed by adding local terms to the
action that are polynomial in boundary curvature invariants \cite{GM}. These
divergences are the de Sitter analogs of those found in the AdS case for
compact boundary geometries of the form of sphere or hyperbolic space with
non-trivial topology \cite{EJM}. For reasons similar to the AdS case, the
linear divergence could be reflective of a UV divergence in the Euclidean
boundary CFT and this need not be fatal to a putative dS/CFT correspondence.

In higher dimensions, by writing the Einstein equations in the
Gauss-Coddacci form we can find the counterterm Lagrangian for any given $d$%
. The result is,

\begin{eqnarray}
\mathcal{L}_{ct} &=&\left( -\frac{d-1}{\ell }+\frac{\ell \Theta \left(
d-3\right) }{2(d-2)}\hat{R}\right) -\frac{\ell ^{3}\Theta \left( d-5\right) 
}{2(d-2)^{2}(d-4)}\left( \hat{R}^{ab}\hat{R}_{ab}-\frac{d}{4(d-1)}\hat{R}
^{2}\right)  \notag \\
&&-\frac{\ell ^{5}\Theta \left( d-7\right) }{(d-2)^{3}(d-4)(d-6)}\left( 
\frac{3d+2}{4(d-1)}\hat{R}\hat{R}^{ab}\hat{R}_{ab}-\frac{d(d+2)}{16(d-1)^{2}}
\hat{R}^{3} \right.  \notag \\
& & \left. -2\hat{R}^{ab}\hat{R}^{cd}\hat{R}_{acbd} - \frac{d}{4(d-1)}\nabla
_{a}\hat{R}\nabla ^{a}\hat{R}+\nabla ^{c} \hat{R}^{ab}\nabla _{c}\hat{R}%
_{ab}\right)  \label{countergen}
\end{eqnarray}
for $d\leq 8$, where the step function $\Theta (x)$ vanishes unless $x>0$.
The associated boundary stress-energy tensor can be obtained by the
variation of the action with respect to the boundary metric.

If the boundary geometry has an isometry generated by a Killing vector $\xi
^{\mu }$, then it is straightforward to show that $T_{ab}\xi ^{b}$ is
divergenceless. We write the boundary metric in the form 
\begin{equation}
\gamma _{ab}d\hat{x}^{a}d\hat{x}^{b}=d\hat{s}^{2}=N_{t}^{2}dt^{2}+\sigma
_{ab}\left( d\varphi ^{a}+N^{a}dt\right) \left( d\varphi ^{b}+N^{b}dt\right)
\label{hmetric}
\end{equation}%
where $\nabla _{\mu }t$ is a spacelike vector field that is the analytic
continuation of a timelike vector field and the $\varphi ^{a}$ are
coordinates describing closed surfaces $\Sigma $. From this it is
straightforward to show that the quantity 
\begin{equation}
\mathfrak{Q}=\oint_{\Sigma }d^{d-1}\varphi \sqrt{\sigma }n^{a}T_{ab}\xi ^{b}
\label{Qcons}
\end{equation}%
is conserved between surfaces of constant $t$, whose unit normal is given by 
$n^{a}$. Physically this would mean that a collection of observers on the
hypersurface whose metric is $\gamma _{ab}$ would all observe the same value
of $\mathfrak{Q}$ provided this surface had an isometry generated by $\xi
^{b}$. If $\partial /\partial t$ is itself a Killing vector, then we can
define 
\begin{equation}
\mathfrak{M}=-\oint_{\Sigma }d^{d-1}\varphi \sqrt{\sigma }%
N_{t}n^{a}n^{b}T_{ab}  \label{Mcons}
\end{equation}%
as the conserved mass associated with the surface $\Sigma $ at any given
point $t$ on the boundary. This quantity changes with the cosmological time $%
\tau $. However a collection of observers that defined a surface $\Sigma $
would find that the value of $\mathfrak{M}$ that they would measure would
not change were they collectively relocated at a different value of $t$ on
the spacelike surface $\mathcal{\partial M}$. Since all asymptotically de
Sitter spacetimes must have an asymptotic isometry generated by $\partial
/\partial t$, there is at least the notion of a conserved total mass $%
\mathfrak{M}$ for the spacetime as computed at future/past infinity.
Similarly the quantity 
\begin{equation}
\mathfrak{J}^{a}=\oint_{\Sigma }d^{d-1}\varphi \sqrt{\sigma }\sigma
^{ab}n^{c}T_{bc}  \label{Jcons}
\end{equation}%
can be regarded as a conserved angular momentum associated with the surface $%
\Sigma $ if the surface has an isometry generated by $\partial /\partial
\phi ^{a}$. Now, by knowing the conserved mass and total action, we can
evaluate the entropy of the gravitational system. We use the relation 
\begin{equation}
S=\lim_{\tau \rightarrow \infty }\left( \beta _{H}\mathfrak{M}-I\ \right)
\label{entr}
\end{equation}%
extending the usual definition to asymptotically dS cases \cite{GM}, which
we justify below. It has been shown for $\left( d+1\right) $--dimensional
SdS space that $S$ is a positive monotonically increasing function of $%
\mathfrak{M}$, and that in $(2+1)$ dimensions $S$ is consistent with the
Cardy formula \cite{bala,GM}, providing suggestive evidence in favour of the
second law in this context.


\section{Path integral formalism in asymptotically dS spacetimes}

\label{sec:pathintegral}

In this section we consider extending the path--integral formulation of
(semi-classical) quantum gravity to the case of asymptotically dS
spacetimes. We will see that this approach provides a justification for the
relation (\ref{entr}), despite the rather unconventional notion of total
mass energy given in eq. (\ref{Mcons}).

We begin by considering the standard path--integral approach, where we want
to take the amplitude to go from a state $\left[ g_{1},\Phi _{1}\right] $ on
a surface $S_{1}$ to state $\left[ g_{2},\Phi _{2}\right] $ on $S_{2}$,
where $g_{i},\Phi _{i}$ represent the metric and matter fields of interest.
Taking the action $I[g,\Phi ]$ to be over all fields on the surfaces $%
S_{1},S_{2}$, this can be represented as 
\begin{equation}
\left\langle g_{2},\Phi _{2},S_{2}|g_{1},\Phi _{1},S_{1}\right\rangle =\int
D \left[ g,\Phi \right] \exp \left( \text{i}I\left[ g,\Phi \right] \right)
\label{PI1}
\end{equation}%
with $D[g,\Phi ]$ a measure of all possible field configurations. For aF and
aAdS, the surfaces are joined by timelike tubes with a finite mean radius,
making the boundary and the interior region contained within compact. By
taking the limit that the larger/smaller mean radii becomes
infinite/vanishes, we get the amplitude for the evolution from $\left[
g_{1},\Phi _{1},S_{1}\right] $ to $\left[ g_{2},\Phi _{2},S_{2}\right] $.

\begin{figure}[tbp]
\begin{center}
\begin{minipage}[c]{.45\textwidth}
         \centering
         \includegraphics[width=\textwidth]{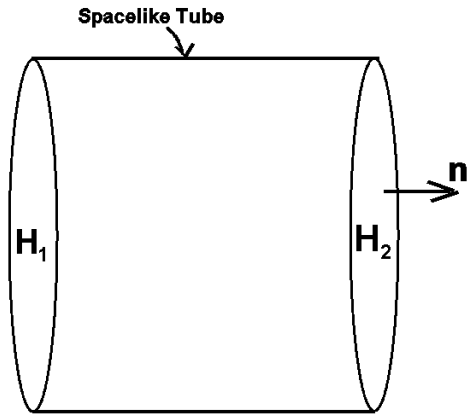}
\end{minipage}
\end{center}
\caption{Histories $H_{1}$ and $H_{2}$,with unit spacelike normal $n$,
joined by spacelike tube}
\label{fig00}
\end{figure}

Asymptotically de Sitter spacetimes, however, must be handled differently.
In $(d+1)$--dimensional adS spacetimes, we replace the surfaces $S_1,S_2$
with histories $H_1,H_2$ that have spacelike unit normals and are surfaces
that form the timelike boundaries of a given spatial region; they therefore
describe particular histories of $d$-dimensional subspaces of the full
spacetime. The amplitude (\ref{PI1}) becomes 
\begin{equation}
\left\langle g_{2},\Phi _{2},H_{2}|g_{1},\Phi _{1},H_{1}\right\rangle =\int
D \left[ g,\Phi \right] \exp \left( \text{i}I\left[ g,\Phi \right] \right)
\label{PI2}
\end{equation}
This now describes an amplitude between differing histories $\left[
g_{1},\Phi_{1}\right] $ and $\left[ g_{2},\Phi_{2}\right]$ instead of
surfaces. The histories $H_1, H_2$ are joined by spacelike tubes at some
initial and final times, so that again the boundary and interior region are
compact (fig. \ref{fig00}), and now, in the limit that the times approach
past/future infinity, we obtain the correlation between complete histories $%
\left[ g_{1},\Phi _{1},H_{1} \right] $ and $\left[ g_{2},\Phi _{2},H_{2}%
\right]$. We arrive at the correlation by summing over all metric and matter
field configurations that interpolate between these two histories. 

By analogy with the aF and aAdS cases, the quantity $\left\langle
g_{2},\Phi_{2}, H_{2}|g_{1}, \Phi_{1}, H_{1}\right\rangle $ depends only on
the hypersurfaces $H_1, H_2$, along with the metrics and matter fields on
these hypersurfaces. It does not depend on any hypersurface lying between
these two.

The correlation between the two histories, found by summing over all
intermediate histories, 
\begin{equation}
\left\langle g_{2},\Phi _{2},H_{2}|g_{1},\Phi _{1},H_{1}\right\rangle
=\sum_{i}\left\langle g_{2},\Phi
_{2},H_{2}|g_{i},\Phi_{i},H_{i}\right\rangle \left\langle
g_{i},\Phi_{i},H_{i}|g_{1},\Phi _{1},H_{1}\right\rangle  \label{PI3}
\end{equation}
shows the necessity of the boundary action (\ref{actbound}), which is
another way of saying that, for an asymptotically dS spacetime, the boundary 
$\partial \mathcal{M}$ will be a union of Euclidean spatial boundaries at
early and late times. Equation (\ref{PI3}) will hold provided 
\begin{equation}
I\left[ g_{12},\Phi \right] =I\left[ g_{1i},\Phi \right] +I\left[
g_{i2},\Phi \right]  \label{action123}
\end{equation}%
where $g_{1i}$ is the metric of the spacetime region between histories $%
H_{1} $ and $H_{i}$ and $g_{i2}$ is the metric of that between histories $%
H_{i}$ and $H_{2}$. The metric $g_{12}$ is thus the metric of the full
spacetime between the two initial and final histories, obtained by joining
the two regions. In general, $g_{1i}$ and $g_{i2}$ will have different
spacelike normal derivatives, yielding delta--function contributions to the
Ricci tensor proportional to the difference between the extrinsic curvatures
of the history $H_{i}$ in $g_{1i}$ and $g_{i2}$. The action $I_{\partial B}$
compensates for these discontinuities, ensuring (\ref{action123}) holds.

We now wish to consider a counter--term action $I_{ct}$ that will apply to
the dS/CFT correspondence. This term appears due to the counterterm
contributions from the boundary quantum CFT \cite{balakraus,CFTref}, found
by analogy with the AdS/CFT correspondence, which suggests the following
relationship 
\begin{eqnarray}
Z_{\text{AdS}}[\gamma ,\Psi _{0}] &=&\int_{[\gamma ,\Psi _{0}]}D\left[ g%
\right] D\left[ \Psi \right] e^{-I\left[ g,\Psi \right] }=\left\langle \exp
\left( \int_{\partial \mathcal{M}{_{d}}}d^{d}x\sqrt{g}\mathcal{O}_{[\gamma
,\Psi _{0}]}\right) \right\rangle  \notag \\
&=&Z_{CFT}[\gamma ,\Psi _{0}]  \label{PAR}
\end{eqnarray}%
between the partition function of the field theory on AdS$_{d+1}$ and the
quantum conformal field theory on its boundary. This correspondence has been
explicated for free and interacting massive scalar fields and free massive
spinor fields (as special cases of interacting scalar-spinor field theory %
\cite{Kav}) along with classical gravity and type IIB string theory \cite%
{Lu,BanksG,Chal}. The counter-term action $I_{ct}$ appears for similar
reasons: the quantum CFT at future/past infinity is expected to have
counterterms whose values can only depend on geometric invariants of these
spacelike surfaces. The counterterm action can be shown to be universal for
both the AdS and dS cases by re-writing the Einstein equations in
Gauss--Codacci form, and then solving them in terms of the extrinsic
curvature and its derivatives to obtain the divergent terms; these will
cancel the divergences in the bulk and boundary actions \cite%
{GM,KrausLarsenSieb}. It can be generated by an algorithmic procedure,
without reference to a background metric, and yields finite values for
conserved quantities that are intrinsic to the spacetime. The result of
employing this procedure in de Sitter spacetime is given by the counterterm
Lagrangian (\ref{countergen}).

Since (\ref{Qcons}) is conserved, a collection of observers on a
hypersurface with metric (\ref{hmetric}) will all have the same $\mathfrak{Q}
$, provided the surface has an isometry generated by $\xi ^{a}$. In other
words, provided $\xi ^{a}$ is a Killing vector on the surface $\Sigma $, $%
\mathfrak{Q}$ is the same on any two histories. Note that $\Sigma $ doesn't
enclose anything, but rather is a boundary of the histories that interpolate
between the initial and final histories. This means that in a sense, $%
\mathfrak{Q}$ is only associated with the boundary, and not with the
histories it bounds. This is analogous to the situation in asymptotically
flat and AdS spacetimes, in which conserved quantities can be associated
with surfaces whose interiors have no isometries \cite{ivan}.

The path integral still remains to be evaluated. It is easier to first
review the path integral approach for aF and aAdS and the relationship
between this approach and gravitational thermodynamics. Consider first a
scalar quantum field $\phi $. We can write the amplitude for going from
states $|{t_{1},\phi _{1}}\rangle $ to $|{t_{2},\phi _{2}}\rangle $ in two
different ways. We can have it as an integral over all possible states
existing between the initial and final states, 
\begin{equation}
\langle {t_{2},\phi _{2}}|{t_{1},\phi _{1}}\rangle =\int_{1}^{2}d[\phi ]e^{%
\text{i}I[\phi ]}  \label{qmamplitude1}
\end{equation}
However, we can also express such an amplitude using the Hamiltonian 
\begin{equation}
\langle {t_{2},\phi _{2}}|{t_{1},\phi _{1}}\rangle =\langle {\phi _{2}}|e^{- 
\text{i}H(t_{2}-t_{1})}|{\phi _{1}}\rangle  \label{qmamplitude2}
\end{equation}
By imposing the periodicity condition $\phi_{1}=\phi _{2}$ for $t_{2}-t_{1}=%
\text{i}\beta $, we sum over $\phi_{1}$ to obtain 
\begin{equation}
\text{Tr}(\exp (-\beta H))=\int d[\phi ]e^{-\hat{I}[\phi ]}.
\label{partition1}
\end{equation}
Note that what has been done here is a Wick rotation of the time coordinate,
giving a Euclidean path integral over the field configurations, where $%
\hat{I}$ is the Euclidean action. Inclusion of gravitational effects can be
carried out as described above, by considering the initial state to include
a metric on a surface $S_{1}$ at time $t_{1}$ evolving to another metric on
a surface $S_{2}$ at time $t_{2}$, yielding the relation (\ref{PI1}).

The left--hand side of (\ref{partition1}) has become the partition function $%
Z$, with temperature $\beta ^{-1}$, for the canonical ensemble for a field.
This connection with standard thermodynamic arguments \cite{Pathria} can be
seen as follows. We start with the canonical distribution 
\begin{equation}
P_{r}\equiv \frac{<n_{r}>}{\mathcal{N}}=\frac{e^{-\beta E_{r}}}{%
\sum_{r}e^{-\beta E_{r}}}  \label{Pr}
\end{equation}%
with $\beta $ determined by considering the average total energy $M$ 
\begin{equation}
M=\frac{\sum_{r}E_{r}e^{-\beta E_{r}}}{\sum_{r}e^{-\beta E_{r}}}=-\frac{%
\partial }{\partial \beta }\ln \left\{ \sum_{r}e^{-\beta E_{r}}\right\} =-%
\frac{\partial }{\partial \beta }\ln Z.  \label{Mnormal}
\end{equation}%
Also, the Helmholtz free energy $W=M-TS$ can be rearranged so that 
\begin{subequations}
\begin{eqnarray}
M=W+TS &=&W-T\left( \frac{\partial W}{\partial T}\right) _{N,V}=-T^{2}\left[ 
\frac{\partial }{\partial T}\left( \frac{W}{T}\right) \right] _{N,V}
\label{Mnormal2} \\
&=&\frac{\partial }{\partial \beta }\left( \beta W\right) .  \label{Mnormal3}
\end{eqnarray}%
Comparing (\ref{Mnormal}) and (\ref{Mnormal3}), we get 
\end{subequations}
\begin{equation}
-\beta W=\ln \left\{ \sum_{r}e^{-\beta E_{r}}\right\} =\ln Z
\label{GDnormal}
\end{equation}%
which can be interpreted as describing the partition function of a
gravitational system at temperature $\beta ^{-1}$ contained in a (spherical)
box of finite radius.

$Z$ can be computed using an analytic continuation of the action in (\ref%
{PI1}) so that the axis normal to the surfaces $S_{1},S_{2}$ is rotated
clockwise by $\frac{\pi }{2}$ radians into the complex plane \cite%
{GibbonsHawking1} (i.e. $t\rightarrow \text{i}T$) in order to obtain a
Euclidean signature. Since the Euclidean action is positive, the path
integral will be convergent, and hence any calculations of desired
quantities (action, entropy, etc.), can be safely carried out without
concern. After the calculations have been done, physical results are
obtained by rotating the system back to a Lorentzian signature.

The action for asymptotically de Sitter spacetimes outside the cosmological
horizon is in general negative, and so the above arguments must be modified
for such cases. Here, one Wick rotates the axis normal to the histories $%
H_1,H_2$ anticlockwise by $\frac{\pi}{2}$ (i.e. $t \rightarrow -\mathit{i}T$%
). This makes the action imaginary, and so $\exp \left( \text{i}I\left[
g,\Phi \right] \right) \longrightarrow \exp \left( +\hat{I}\left[ g,\Phi %
\right] \right) $, giving convergent path integral 
\begin{equation}
Z^{\prime }=\int e^{+\hat{I}}  \label{Partitionaction}
\end{equation}
since $\hat{I}<0$. Furthermore, since we want a converging partition
function, we must change (\ref{Mnormal}) to 
\begin{equation}
M=+\frac{\partial }{\partial \beta }\ln \left\{ \sum_{r}e^{+\beta
E_{r}}\right\} =+\frac{\partial }{\partial \beta }\ln Z^{\prime }.
\label{Mout}
\end{equation}
Now comparing (\ref{Mout}) with (\ref{Mnormal3}) (since (\ref{Mnormal2},\ref%
{Mnormal3}) won't change) we will obtain 
\begin{equation}
+\beta W=\ln \left\{ e^{+\beta E_{r}}\right\} =\ln Z^{\prime }.
\label{GDout1}
\end{equation}
In the semi-classical approximation this will lead to $\ln Z^{\prime
}=+I_{cl}$. Substituting this and (\ref{Mnormal2}) into (\ref{GDout1}), 
\begin{eqnarray}
\beta \left( M-TS\right) &=&+I_{cl}  \notag \\
\beta M-S &=&I_{cl}  \notag \\
S &=&\beta M-I_{cl}.  \label{GDoutfinal}
\end{eqnarray}

As before, the presumed physical interpretation of the results is then
obtained by rotation back to a Lorentzian signature at the end of the
calculation. Here, however, we are working outside the horizon in an
asymptotically dS spacetime, where the signature near past and future
infinity becomes $(+,-,+,+,\ldots)$, giving the spacelike boundary tubes a
Euclidean signature. There are thus two ways to proceed.

One approach is suggested by the form of $\partial _{t}$ at future infinity,
which is asymptotically a spacelike Killing vector. This suggests \cite{GM}
that the Wick rotation employed above is used merely to establish the
relation (\ref{GDoutfinal}), with no need to analytically continue either
the time or any other values such as rotation or NUT parameters. This would
mean that one can evaluate the action at future infinity, imposing the
periodicity in $t$ consistent with regularity at the cosmological horizon.
Since no quantities are analytically continued, we refer to this approach as
the R--approach, and demonstrate it in section \ref{sec:examples}.

Another approach is to use the Wick rotation and calculate all quantities in
a manner analogous to the approaches used for aF or aAdS spacetimes ---
since there is an analytic continuation used here, we refer to this as the
C--approach. This will involve rotation of both the time (spacelike outside
the horizon; $t\rightarrow \text{i}T$) and any rotation or NUT charge
parameters involved in the metric. This will give a signature $%
(-,-,+,+,\ldots )$, which will cause the calculated action to be negative,
giving a negative definite energy. The argument for this approach is that
the important factor is the convergence of the path integral and partition
function, as opposed to simply obtaining a Euclidean signature. The new time 
$T$ is periodically identified with the period $\beta $ to ensure the
absence of conical singularities. We use this C--approach in section \ref%
{sec:examplesC}.

In fact, it can be shown that these two approaches are completely equivalent
modulo analytic continuation \cite{RCstelea}. We can start from the
R--approach results and derive by consistent analytic continuations all
results from the C--approach. Alternatively, there are no obstacles in
beginning with the C--approach and deriving from this method the respective
R--approach results. We shall employ both approaches in this paper, keeping
in mind that they are related by analytic continuation.

The main metric of interest in this paper is the Taub--NUT metric. This
metric possesses a spurious singularity known as a Misner--string
singularity, analogous to the singularity that arises in electromagnetic
theory in considering the Dirac monopole. In order to avoid this
singularity, an additional periodicity constraint in $t$ must be imposed.
This periodicity is incorporated with the periodicity $\beta $, yielding a
consistency criterion relating the mass and NUT parameters, the two
solutions of which produce generalizations of asymptotically flat Taub-Bolt
space to the asymptotically de Sitter case \cite{dsnutshort}. These
solutions are classified based on the dimension of the fixed point set of
the Killing vector $\partial _{t}$ that generates the $U(1)$ isometry group.
The solution is a Bolt solution if the fixed point set is $(d-1)$%
--dimensional, and a NUT solution if it is less than this. The R--approach
only produces a Bolt \cite{CGM2}; however, the C--approach will produce both
a NUT and a Bolt solution, similar to the AdS--NUT case. It is as always
important to note that both of these versions --- the C--approach and the
R--approach versions --- are solutions of the Einstein equations of motion.

\section{The $\mathbf{N}$--bound}

\label{sec:N-bound}

The Bousso $\mathbf{N}$--bound is a stronger version, and a non--trivial
prediction, of the Banks $\Lambda -\mathbf{N}$ proposal. Banks proposed that
the cosmological constant, as an input parameter, should be determined by
the inverse of the number of degrees of freedom of the fundamental theory %
\cite{Banks}. As a result of this proposal, a four dimensional universe with
positive cosmological constant $\Lambda $ tends to evolve to empty de Sitter
spacetime, and has at most $\mathbf{N}=\frac{3\pi }{\Lambda }=\pi \ell ^{2}=%
\frac{1}{4}A_{c.h.}$ degrees of freedom equal to the finite entropy of the
empty de Sitter spacetime, where $A_{c.h.}$ is the area of the cosmological
horizon. So, the $\Lambda -\mathbf{N}$ proposal says that a universe with
positive cosmological constant cannot have entropy greater than that of pure
de Sitter spacetime. This prediction was called the $\mathbf{N}$--bound and
its proof was presented by Bousso \cite{bousso}. The proof is based on a
combination of the covariant entropy bound, the D--bound, the concept of
causal diamonds and Bekenstein's generalized second law of thermodynamics %
\cite{BEK}.

The covariant entropy bound bounds the entropy on certain light-sheets. It
states that the entropy on any light-sheet is less than or equal to $\frac{A%
}{4}$, where $A$ is the surface of the light-sheet. This bound generalizes a
proposal by Fischler and Susskind \cite{SUS} and is thought to have its
origin in the holographic principle, first formulated by 't Hooft \cite%
{tHooft} and Susskind \cite{SUS2}. The application of the covariant entropy
bound to the past light-cone of an observer was proposed by Banks and a more
stringent covariant bound was obtained for light sheets that do not
terminate on caustics by Flanagan \textit{et al. }\cite{Flanagan}.

The D--bound on matter entropy of the asymptotically de Sitter spacetimes is
a generalization of Bekenstein's bound on the entropy of the finite systems
in asymptotically flat spacetimes \cite{Beken}. Bekenstein's bound can be
written as $S_{m}\leq 2\pi mR$, where $2m$ is the gravitational radius of
the system and $R$ is the circumscribing radius of the system. The D--bound
states that the matter entropy is less than the difference between $\mathbf{N%
}$ and a quarter of the area of the de Sitter cosmological horizon. For
dilute, spherically symmetric systems in de Sitter space, the D--bound takes
precisely the form of Bekenstein's bound, despite the significant deviation
from flat space \cite{Bousso3}.

The explicit $\mathbf{N}$--bound conjecture states that in any universe with
a positive cosmological constant, as well as additional matter that may well
dominate at all times, the observable entropy is bounded by $\mathbf{N}= 
\frac{3\pi }{\Lambda }$.

\begin{figure}[tbp]
\begin{center}
\begin{minipage}[c]{.45\textwidth}
         \centering
         \includegraphics[width=\textwidth]{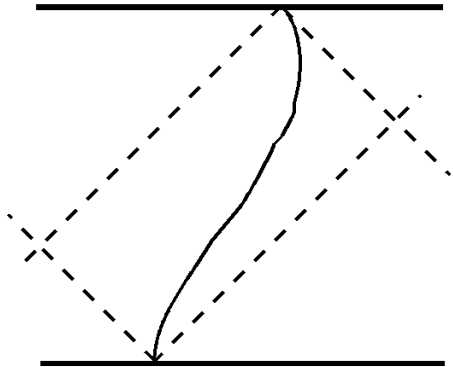}
\end{minipage}
\end{center}
\caption{Causal diamond of an observer's world line between initial and
final time slices}
\label{fig000}
\end{figure}

The observable entropy includes both matter and horizon entropy, but
excludes entropy that cannot be observed in a causal experiment. In finding
the observable entropy, one should restrict attention to the causal diamond
of an observer which is the spacetime region that can be both influenced and
seen by an observer. So, the observable entropy lies in a region bounded by
the past and future light cones from the endpoints of the observer's world
line (Fig. \ref{fig000}). 

We note that $\mathbf{N}$ is the Bekenstein--Hawking entropy of empty de
Sitter space and the bound becomes trivial in the limit of vanishing
cosmological constant.

The most profound implication of the $\mathbf{N}$--bound is that a quantum
gravity theory with an infinite number of degrees of freedom, such as
M--theory, cannot describe correctly spacetimes with a positive cosmological
constant. In fact, this is in good agreement with the other fact that no
stable de Sitter solutions are yet known in M--theory.

\section{Maximal Mass Conjecture}

\label{sec:maxmass}

Balasubramanian, de Boer and Minic \cite{bala} proposed a method for
computing the boundary stress tensor, mass and the conserved charges of
asymptotically dS spacetimes from data at early or late time infinity. The
method is analogous to the Brown-York prescription in asymptotically AdS
spacetimes \cite{balakraus,brown,BCM,ivan}, suggesting a holographic duality
similar to the AdS/CFT correspondence.

In this context, the conserved charge associated with the Killing vector $%
\partial /\partial t$\ --- now spacelike outside of the cosmological horizon
--- is interpreted as the conserved mass of the spacetime. Using this
definition, Balasubramanian \textit{et al. }\cite{bala} found the conserved
masses for the four and five dimensional Schwarzschild--dS spacetimes and
three dimensional Kerr--dS spacetime and were led to the conjecture that 
\textit{any asymptotically dS spacetime with mass greater than dS has a
cosmological singularity}. This is what we shall refer to as the maximal
mass conjecture. We interpret the cosmological singularities in the maximal
mass conjecture to imply that scalar Riemann curvature invariants will
diverge to form timelike regions of geodesic incompleteness whenever the
conserved mass of a spacetime becomes larger than the zero value of pure dS.
As stated before, the conjecture is in need of a proof. Since the mass
formula is constructed in terms of the extrinsic curvature of the spacetime
boundary, standard techniques using the Raychaudhuri equation and focussing
theorems may be applicable in this direction.

The conjecture has been verified for topological dS solutions and their
dilatonic variants \cite{cai}, the Schwarzschild-de Sitter (SdS) black hole
up to dimension nine \cite{GM} and Reissner-Nordstorm-de Sitter black hole %
\cite{AMR}.

As stated before, by carrying out a procedure analogous to that in the AdS
case \cite{balakraus,BCM}, one can compute the boundary stress tensor of the
spacetime, and from this calculate a conserved charge interpreted as the
mass of the asymptotically dS spacetime. Another interesting point is that
the trace of boundary stress tensor can be employed in the determination of
the dual RG equation \cite{Stro}, by assuming dS/CFT holography. So, RG
evolution of the dual field theory is time evolution in an expanding
universe. In other words, the evolution of the central charge in the dual
field theory is related to the changing number of accessible degrees of
freedom in spacetime.

\section{Cosmological Schwarzschild--Taub/NUT-Kerr Spacetimes}

\bigskip \label{sec:cosmo-spaceimes}

In this section, we consider the best known asymptotically de Sitter
spacetimes outside the cosmological horizon. One reason for doing this is
that by choosing this region of spacetime, extended from the cosmological
horizon to the boundary at late time, we avoid the other boundary at early
time, simplifying the calculation.

We begin by considering the $d+1$ Schwarzschild-dS spacetime with the metric 
\begin{equation}
ds^{2}=-N(r)dt^{2}+\frac{dr^{2}}{N(r)}+r^{2}d\hat{\Omega}_{d-1}^{2}
\label{Sdsmetric}
\end{equation}%
where 
\begin{equation}
N(r)=1-\frac{2m}{r^{d-2}}-\frac{r^{2}}{\ell ^{2}}  \label{Sdslapse}
\end{equation}%
and $d\hat{\Omega}_{d-1}^{2}$ denotes the metric on the unit sphere $S^{d-1}$%
. For a mass parameter $m$ with $0<m<m_{N}$, where 
\begin{equation}
m_{N}=\frac{\ell ^{d-2}}{d}\left( \frac{d-2}{d}\right) ^{\frac{d-2}{2}}
\label{naraimass}
\end{equation}%
we have a black hole in dS spacetime with event horizon at $r=r_{H}$ and
cosmological horizon at $r=r_{C}>r_{H}$. The event and cosmological horizons
are located at $N(r_{H})=N(r_{C})=0.$ When $m=m_{N},$ the event horizon
coincides with the cosmological horizon and one gets the Nariai solution.
For $m>m_{N}$, the metric (\ref{Sdsmetric}) describes a naked singularity in
an asymptotically dS spacetime. So demanding the absence of naked
singularities yields an upper limit to the mass of the SdS black hole.
Outside the cosmological horizon, $N(r)<0$, \ so we set $r=\tau $ and
rewrite the metric as 
\begin{equation}
ds^{2}=-f(\tau )d\tau ^{2}+\frac{dt^{2}}{f(\tau )}+\tau ^{2}d\tilde{\Omega}
_{d-1}^{2}  \label{Sdsmet2}
\end{equation}
where 
\begin{equation}
f(\tau )=\left( \frac{\tau ^{2}}{\ell ^{2}}+\frac{2m}{\tau ^{d-2}}-1\right)
^{-1}  \label{flapse}
\end{equation}

The Lorentzian Kerr--dS geometry is given by 
\begin{eqnarray}
ds^{2} & = & -\frac{\Delta _{L}(r)}{\Xi _{L}^{2}\rho _{L}^{2}}(dt-a\sin
^{2}\theta d\phi )^{2} + \frac{\Theta _{L}(\theta )\sin ^{2}\theta }{\Xi
_{L}^{2}\rho _{L}^{2}}[adt-(r^{2}+a^{2})d\phi ]^{2}  \notag \\
& & + \frac{\rho _{L}^{2}dr^{2}}{\Delta _{L}(r)}+\frac{\rho _{L}^{2}d\theta
^{2}}{\Theta _{L}(\theta )}  \label{KdSLmetr}
\end{eqnarray}
where 
\begin{equation}
\begin{array}{rcl}
\rho _{L}^{2} & = & r^{2}+a^{2}\cos ^{2}\theta \\ 
\Delta _{L}(r) & = & -\frac{r^{2}(r^{2}+a^{2})}{\ell ^{2}}+r^{2}-2mr+a^{2}
\\ 
\Theta _{L}(\theta ) & = & 1+\frac{a^{2}}{\ell ^{2}}\cos ^{2}\theta \\ 
\Xi _{L} & = & 1+\frac{a^{2}}{\ell ^{2}}.%
\end{array}
\label{KdSLfuncs}
\end{equation}
The event horizons of the spacetime are given by the singularities of the
metric function, which are the real roots of $\Delta_{L}(r)=0$. The
Lorentzian section is restricted to $\Delta _{L}(r)\geq 0$. Hence the
horizons are determined by the solutions of the equation 
\begin{equation}
r_{H}^{4}-r_{H}^{2}(\ell ^{2}-a^{2})+2m\ell ^{2}r_{H}-\ell ^{2}a^{2}=0.
\label{horizons}
\end{equation}
In the limit $\ell \rightarrow \infty $, equation (\ref{horizons}) yields
the well known location of the Kerr black hole horizon 
\begin{equation}
r_{H}=m+\sqrt{m^{2}-a^{2}}.  \label{Kerrhorizon}
\end{equation}
For small rotation parameter ($a\rightarrow 0$), equation (\ref{horizons})
reduces to 
\begin{equation}
\frac{r_{H}^{3}}{\ell ^{2}}-r_{H}+2m=0  \label{horizonsSdS}
\end{equation}
which gives us the location of Schwarzschild--dS event horizon $r_{H}$ and
cosmological horizon $r_{C}$. In this case, for mass parameters $m$ with $
0<m<m_{N}$, where 
\begin{equation}
m_{N}=\frac{\ell }{3\sqrt{3}}  \label{schnaraimass}
\end{equation}
we have a black hole in dS spacetime with event horizon at $r=r_{H}$ and
cosmological horizon at $r=r_{C}>r_{H}$. When $m=m_{N}$, the event horizon
coincides with the cosmological horizon $r_{C}=r_{H}$ $=\frac{\ell }{\sqrt{3}
}$ and one gets the small rotating Nariai solution.

The other extreme case is when $a\rightarrow \infty $, where equation (\ref%
{horizons}) gives 
\begin{equation}
r_{H}=\ell  \label{horizonsainf}
\end{equation}
for the horizon. For a fixed value of $a$, from the horizon equation (\ref%
{horizons}), we can find the following equations for the extremum of the
horizon radius 
\begin{equation}
2r_{H}+m\mp \sqrt{m^{2}+8mr_{H}}-2r_{H}^{3}/\ell ^{2}=0  \label{exthor}
\end{equation}
where for a fixed $\ell$, the upper branch has a maximum at $r_{+}=(\frac{3 
}{4}+\frac{\sqrt{3}}{2})m$. The corresponding critical cosmological
parameter is $\ell _{+}=\frac{m}{4}(2\sqrt{3}+3)^{3/2}$.

In general for the metric (\ref{KdSLmetr}), the rotating Nariai solution has
an event horizon (coinciding with the cosmological horizon) at 
\begin{equation}
r_{H}=\frac{3m+\sqrt{9m^{2}-8a^{2}(1-\frac{a^{2}}{\ell ^{2}})}}{2(1-\frac{
a^{2}}{\ell ^{2}})}  \label{rHNaraiKdS}
\end{equation}%
which reduces to $r_{H}$ $=\frac{\ell }{\sqrt{3}}$ when $a=0$.

Outside of the cosmological horizon, the Kerr--dS metric function $%
\Delta_{L}(r)$ is negative, so we set $r=\tau $ and rewrite the line element
in the form 
\begin{equation}
ds^{2} = -\frac{\rho ^{2}d\tau ^{2}}{\Delta (\tau )}+\frac{\Delta (\tau )}{%
\Xi ^{2}\rho ^{2}}(dt-a\sin ^{2}\theta d\phi )^{2} +\frac{\Theta (\theta
)\sin ^{2}\theta }{\Xi ^{2}\rho ^{2}}[adt-(\tau ^{2}+a^{2})d\phi ]^{2}+\frac{%
\rho ^{2}d\theta ^{2}}{\Theta (\theta )}  \label{KdSoutmetr}
\end{equation}
where 
\begin{equation}
\begin{array}{rcl}
\rho ^{2} & = & \tau ^{2}+a^{2}\cos ^{2}\theta \\ 
\Delta (\tau ) & = & \frac{\tau ^{2}(\tau ^{2}+a^{2})}{\ell ^{2}}-\tau
^{2}+2m\tau -a^{2} \\ 
\Theta (\theta ) & = & \Theta _{L}(\theta ) \\ 
\Xi & = & \Xi _{L}.%
\end{array}
\label{KdSoutfuncs}
\end{equation}
The angular velocity of the horizon is given by 
\begin{equation}
\Omega _{H}=\left. -\frac{g_{t\phi }}{g_{\phi \phi }}\right|_{\tau =\tau
_{c}}=\frac{a}{\tau_{c}^{2}+a^{2}}  \label{KdSangvel}
\end{equation}
where $\tau _{c}$ is the cosmological horizon $(\Delta (\tau _{c})=0$ and $
\Delta (\tau >\tau _{c})>0)$.

As the last example, we consider the general form of the NUT--charged dS
spacetime in ($d+1$) dimensions, outside the cosmological horizon. The
metric is given by 
\begin{equation}
ds^{2}=V(\tau )\left( dt+nA\right) ^{2}-\frac{d\tau ^{2}}{V(\tau )}+(\tau
^{2}+n^{2})d\Gamma ^{2}  \label{TNDSgeneral}
\end{equation}
where $~d=2k+1$ and $V(\tau )$ is given by the general formula 
\begin{equation}
V(\tau )=\frac{2m\tau }{(\tau ^{2}+n^{2})^{k}}-\frac{\tau }{(\tau
^{2}+n^{2})^{k}}\int_{\tau }ds\left[ \frac{(s^{2}+n^{2})^{k}}{s^{2}}-\frac{
(2k+1)}{\ell ^{2}}\frac{(s^{2}+n^{2})^{k+1}}{s^{2}}\right]  \label{FtBoltgen}
\end{equation}
with $n$ the non--vanishing NUT charge and $\Lambda =\frac{d(d-1)}{2\ell ^{2}%
} $.

The one--form $A$ is a function of the coordinates $(\vartheta_{1},\phi_{1},
\cdot \cdot \cdot ,\vartheta_{k},\phi_{k})$ of the non-vanishing compact
base space of positive curvature (with metric $d\Gamma ^{2}$). The
coordinate $t$ parameterizes a circle $S^{1}$ Hopf--fibred over this space;
it must have periodicity $\frac{2(d+1)\pi \left| n\right| }{q}$ to avoid
conical singularities, where $q$ is a positive integer. The geometry of a
constant--$\tau $ surface is that of a Hopf fibration of $S^{1}$ over the
base space, which is a well defined spacelike hypersurface in spacetime
where $V(\tau )>0$ outside of the past/future cosmological horizons. The
spacelike Killing vector $\partial /\partial t$\ has a fixed point set where 
$V(\tau _{c})=0$\ whose topology is the same as that of the base space.

The general form of the base space is a combination of products of $S^{2}$
and $\mathbb{CP}^{2}$, i.e. $\otimes _{i=1}^{s}S^{2}\otimes _{j=1}^{c} 
\mathbb{CP}^{2}$ such that $s+2c=k$. The metric of $\mathbb{CP}^{2}$ has the
general form 
\begin{equation}
d\Sigma ^{2}=\frac{1}{(1+\frac{u^{2}}{6})^{2}} \left\{du^{2}+\frac{u^{2}}{4}
(d\psi +\cos \theta d\phi )^{2} \right\} + \frac{u^{2}}{4(1+\frac{u^{2}}{6})}
(d\theta^{2}+\sin ^{2}\theta d\phi ^{2})  \label{CP2}
\end{equation}
for which the one--form $A$ is 
\begin{equation}
A=\frac{u^{2}}{2(1+\frac{u^{2}}{6})}(d\psi +\cos \theta d\phi )  \label{A}
\end{equation}
whereas 
\begin{equation}
A=2\cos \theta d\phi  \label{Asph}
\end{equation}
if the base space is a 2-sphere with metric $d\Omega ^{2}=d\theta ^{2}+\sin
^{2}\theta d\phi ^{2}$. For the general form $\otimes _{i=1}^{s}S^{2}\otimes
_{j=1}^{c}\mathbb{CP}^{2}$ the one-form $A$ is a linear combination of
metrics of the forms (\ref{A}) and (\ref{Asph}).

\begin{figure}[tbp]
\centering       
\begin{minipage}[c]{.55\textwidth}
         \centering
         \includegraphics[width=\textwidth]{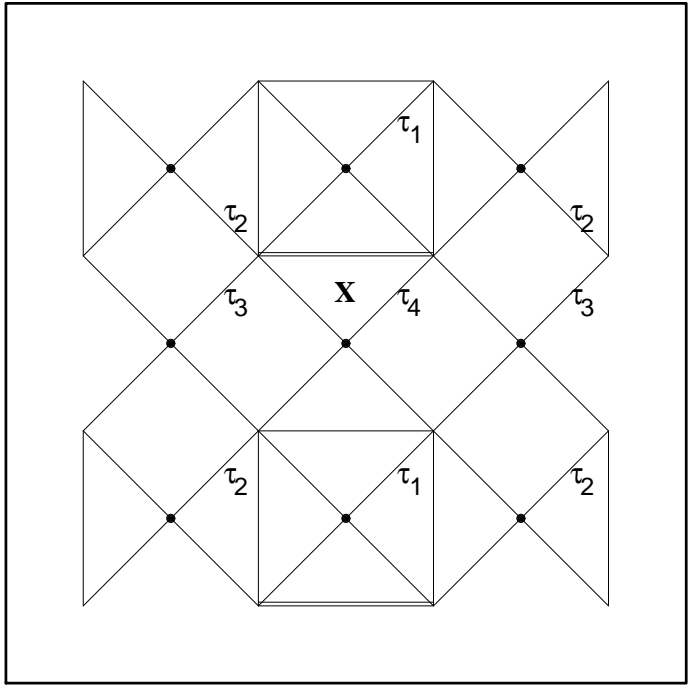}
\end{minipage}
\caption{The Penrose diagram of the TBdS spacetime. We denote the roots of $%
V $ by the increasing sequence $\protect\tau _{1}<0<\protect\tau _{2}< 
\protect\tau _{3}<\protect\tau _{4}=\protect\tau _{c}$. The vertical and
horizontal lines are the $\protect\tau =0$ and the $\protect\tau =-\infty $
slices of the spacetime, respectively. The double line denotes $\protect\tau %
=+\infty $ and the solid black dots denote the quasiregular singularities of
the spacetime. Our calculation is performed outside the cosmological
horizon, located within the triangle denoted by ``X''. }
\label{penrose}
\end{figure}

The causal structure of TNdS spacetime can be understood by looking at a
typical Penrose diagram (Fig. \ref{penrose}). For simplicity, we consider a
four-dimensional TNdS with an $S^{2}$ base space. We denote the roots of $
V(\tau )$ by the increasing sequence $\tau _{1}<0<\tau _{2}<\tau _{3}<\tau
_{4}=\tau_{c}$. The vertical and horizontal lines are the $\tau =0$ and the
past infinity $\tau =-\infty $ slices of the spacetime, respectively and the
double line denotes the future infinity $\tau =+\infty $. The solid black
dots denote the quasiregular singularities. The region that is outside the
cosmological horizon is located inside the triangle denoted by ``X''. 

Quasiregular singularities are the end points of incomplete and inextensible
geodesics which spiral infinitely around a topologically closed spatial
dimension. Moreover the world lines of observers approaching these points
come to an end in a finite proper time \cite{Konkowski}. They are the
mildest kinds of singularity in that the Riemann tensor and its derivatives
remain finite in all parallelly propagated\ orthonormal frames. Consequently
observers do not see any divergences in physical quantities when they
approach a quasiregular singularity. The flat Kasner spacetimes on the
manifolds $\mathbb{R}\otimes T^{3}$ or $\mathbb{R}^{3}\otimes S^{1}$,
Taub-NUT spacetimes and Moncrief spacetimes are some typical spacetimes with
quasiregular singularities.

We consider these quasiregular singularities to be quite different from the
cosmological singularities referred to in the maximal mass conjecture. This
conjecture states that a timelike singularity will be present for any adS
spacetime whose conserved mass (\ref{Mcons}) is positive (i.e. larger than
the zero value of de Sitter spacetime). Using Schwarzschild-de Sitter
spacetime as a paradigmatic example, it is straightforward to show that
scalar Riemann curvature invariants will diverge for $\mathfrak{M}>0$ \cite%
{Shiro,bala,GM}, yielding a timelike boundary to the manifold upon their
excision. Note that such curvature invariants diverge in certain regions
even if $\mathfrak{M}<0$; however observers at future infinity cannot
actively probe such regions. We therefore interpret the cosmological
singularities in the maximal mass conjecture of Balasubramanian \textit{et
al. }\cite{bala} to imply that scalar Riemann curvature invariants will
diverge to form timelike regions of geodesic incompleteness whenever the
conserved mass of a spacetime becomes positive (i.e. larger than the zero
value of pure dS). By this definition quasiregular singularities are clearly
not cosmological singularities, and vice--versa.

We pause here to point out some differences between the four--dimensional
Taub--Bolt and Taub--NUT spaces. For simplicity, we consider the case of
spacetimes with zero cosmological constant. The Euclidean geometry of these
spacetimes is given by the metric 
\begin{equation}
ds^{2}=v_{4}^{-1}(r)dr^{2}+v_{4}(r)\left( d\psi +2n\cos \theta d\phi \right)
^{2}+(r^{2}-n^{2})\left( d\theta ^{2}+\sin ^{2}\theta d\phi ^{2}\right)
\label{dstb4}
\end{equation}
where the function $v_{4}(r)$ is given by 
\begin{equation}
v_{4}(r)=\frac{r^{2}+n^{2}-2mr}{r^{2}-n^{2}}.  \label{f4}
\end{equation}

To have a Bolt solution, which means the fixed point set of the Killing
vector $\partial /\partial \psi $ is a two-dimensional sphere, we should fix 
$m=\frac{5}{4}n,$ such that $v_{4}$ reduces to 
\begin{equation}
v_{4,Bolt}(r)=\frac{(r-2n)(2r-n)}{2(r^{2}-n^{2})}.  \label{frtb4}
\end{equation}%
The metric function $v_{4}$ vanishes at $r=r_{b}=2n>n$ and in this case, the
fixed point set of the Killing vector $\partial /\partial \psi $ is a
two-dimensional sphere with radius $\sqrt{3}n.$

On the other hand, if we fix the mass parameter to be $m=n,$ then we have a
NUT solution at $r=r_{n}=n$, where the fixed point set of the Killing vector 
$\partial /\partial \psi $ is zero-dimensional. In this case, the function $%
v_{4}(r)$ reduces to 
\begin{equation}
v_{4,NUT}(r)={\frac{r-n}{r+n}.}  \label{frtn4}
\end{equation}%
We note that in both cases, 
\begin{equation}
\begin{array}{c}
\left. v_{_{4,Bolt}}(r)\right| _{r=r_{b}}=\left. {\textstyle}%
v_{_{4,NUT}}(r)\right| _{r=r_{n}}=0 \\ 
\left. {\textstyle\frac{d}{dr}}({\textstyle}v_{_{4,Bolt}}(r))\right|
_{r=r_{b}}=\left. {\textstyle\frac{d}{dr}}({\textstyle}v_{4,NUT}(r))\right|
_{r=r_{n}}={\textstyle\frac{1}{2n}.}%
\end{array}
\label{vvp}
\end{equation}%
We also note that the first condition in (\ref{vvp}) is a necessary
condition for the existence of the fixed point set of the Killing vector
(two-dimensional set for Bolt solution and zero dimensional for NUT
solution). The second condition in (\ref{vvp}) is another necessary
condition to avoid a conical singularity at the location of the Bolt ($%
r=r_{b}$) or NUT ($r=r_{n}$). In general, to avoid the conical singularity,
we should have 
\begin{equation}
\Delta \psi {\frac{d}{dr}}v_{4}(r)=4\pi  \label{con}
\end{equation}%
where the derivative is calculated on the location of the Bolt or NUT charge.

Moreover, to avoid the Dirac--Misner string, the period of the coordinate $
\psi $ should be related to the period of $\phi $ by $\Delta \psi =4n\Delta
\phi $. On the other hand, to avoid conical singularities at the poles of
the sphere $(\theta ,\phi )$, we should have $\Delta \phi =2\pi $ and
consequently $\Delta \psi =8\pi n.$ Combining this last relation with the
eq. (\ref{con}), we find the second condition in (\ref{vvp}).

\section{Examples --- R approach}

\label{sec:examples}

In this section, we present in full detail the calculation of the conserved
charges and entropy for the asymptotically dS spacetimes with NUT charge in
different dimensionality, noting violations of the $\mathbf{N}$--bound and
maximal mass conjectures in some cases.

\subsection{Four dimensional case}

The metric of the NUT-charged dS spacetimes is of the form 
\begin{equation}
ds^{2}=V(\tau )(dt+2n\cos \theta d\phi )^{2}-\frac{d\tau ^{2}}{V(\tau )}%
+(\tau ^{2}+n^{2})(d\theta ^{2}+\sin ^{2}\theta d\phi ^{2})  \label{TBDS}
\end{equation}%
where 
\begin{equation}
V(\tau )=\frac{\tau ^{4}+\left( 6n^{2}-\ell ^{2}\right) \tau ^{2}+2m\ell
^{2}\tau -n^{2}\left( 3n^{2}-\ell ^{2}\right) }{\left( \tau
^{2}+n^{2}\right) \ell ^{2}}  \label{VV}
\end{equation}%
with $n$ the nonvanishing NUT charge. The spacelike Killing vector $\partial
/\partial t$\ has a fixed point set where $V(\tau _{0})=0$ whose topology is
that of a 2--sphere. Since $\frac{\partial }{\partial \phi }$ is a Killing
vector, for any constant $\phi $-slice near the horizon $\tau =\tau _{0}$
additional conical singularities will be introduced in the $(t,\tau )$
Euclidean section unless $t$ has period $4\pi /\left| V^{\prime }\left( \tau
_{0}\right) \right| $. This period must equal $\left| \frac{8\pi n}{q}%
\right| $, which forces $\tau _{0}=\tau ^{\pm }$ where $q$ is an integer and 
\begin{equation}
\tau ^{\pm }=\frac{q\ell ^{2}\pm \sqrt{q^{2}\ell ^{4}-144n^{4}+48n^{2}\ell
^{2}}}{12n}  \label{taus}
\end{equation}%
yielding two distinct extensions of the Taub-Bolt-de Sitter spacetime (TB$%
^{\pm }$). The respective mass parameters are 
\begin{equation}
m^{\pm }=-\frac{q^{3}\ell ^{6}\pm (288n^{4}-24\ell ^{2}n^{2}+q^{2}\ell ^{4})%
\sqrt{q^{2}\ell ^{4}-144n^{4}+48n^{2}\ell ^{2}}}{864n^{3}\ell ^{2}}
\label{massesonn}
\end{equation}%
where 
\begin{equation}
\left| n\right| \leq \frac{1}{6}\ell \sqrt{6+3\sqrt{4+q^{2}}}  \label{nmax}
\end{equation}%
so that $\tau ^{\pm }$ are both real. Without loss of generality we can take 
$n>0$; results for $n<0$ can be obtained by reversing the signs of $t$ and $%
\phi $. Note that TB$^{-}$\ does not exist for $\left| n\right|
<n_{c}=.2658\,\ell $\ since $V^{-}(\tau )$ then develops two additional
larger real roots, and the periodicity condition cannot be satisfied.

The spacetime (\ref{TBDS}) is free of scalar curvature singularities --- the
invariants $R_{\mu \nu \rho \lambda }R^{\mu \nu \rho \lambda }$\ and $\sqrt{
-g}R_{\mu \nu }^{\quad \,\alpha \beta }\epsilon _{\alpha \beta \rho \lambda
}R^{\mu \nu \rho \lambda }$ are finite everywhere. The only singularities
are the quasiregular singularities noted above.

\begin{figure}[tbp]
\begin{center}
\begin{minipage}[c]{.45\textwidth}
         \centering
         \includegraphics[width=\textwidth]{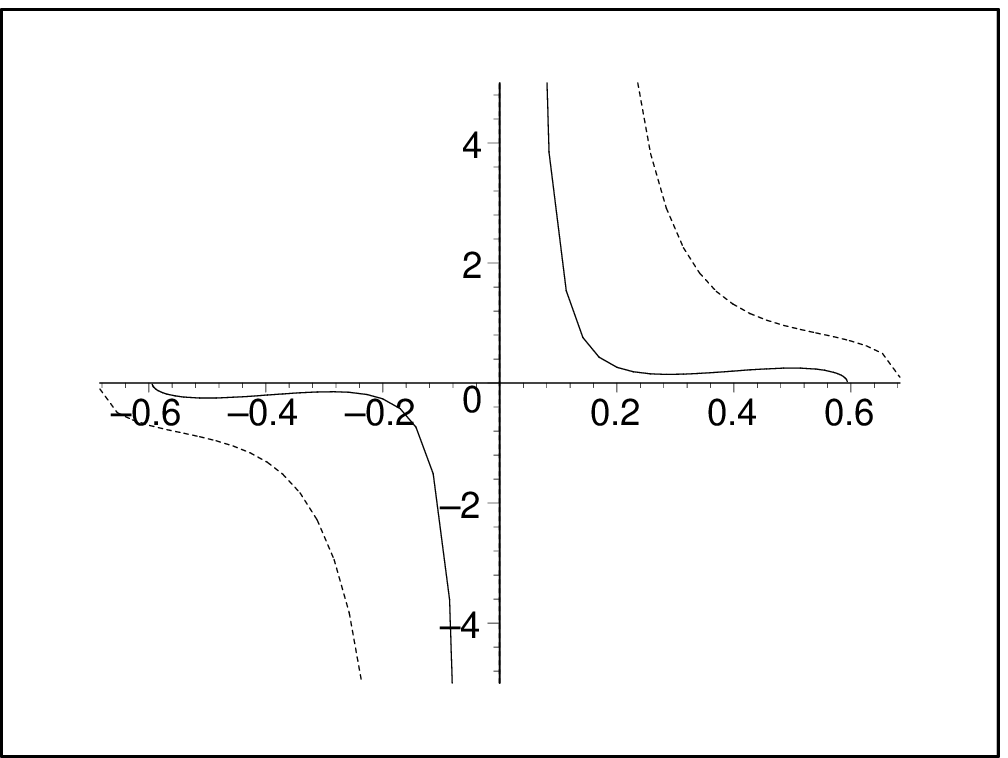}
         \label{MPLUSK13BW}
     \end{minipage}
\begin{minipage}[c]{.45\textwidth}
         \centering
         \includegraphics[width=\textwidth]{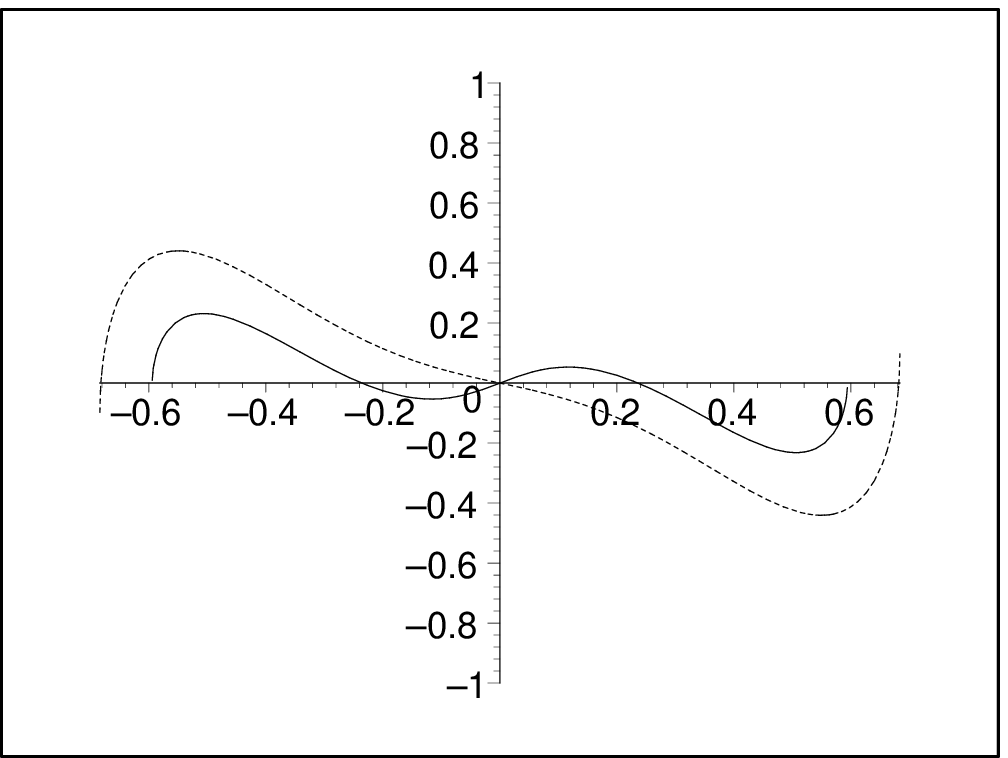}
         \label{MMINUSK13BW}
     \end{minipage}
\end{center}
\caption{Left/right: Total mass $\mathfrak{M}^{+/-}\mathfrak{/}\mathbb{\ell }
$ of the TB$^{+/-}$ solution versus $n/\ell $ at future infinity, for $q=1$
(solid) and $q=3$ (dotted)$.$ The spacetimes TB$^{-}$ with $q=1$ and $q=3$
exist for $\left| n\right| >n_{c}\simeq .2658$ and $\left| n\right|
>n_{c}=.1879$ respectively$.$ }
\label{fig1}
\end{figure}

For the conserved mass (\ref{Mcons}), we obtain 
\begin{equation}
\mathfrak{M}=-m+\frac{105n^{4}-30n^{2}\ell ^{2}+\ell ^{4}}{8\ell ^{2}\tau }+%
\mathcal{O}\left( \frac{1}{\tau ^{2}}\right)  \label{totmass}
\end{equation}%
near future infinity, which for $n=0$ reduces exactly to the total mass of
the $d=4$ Schwarzschild--dS black hole \cite{GM}. Note that although for the 
$d=4$ Schwarzschild--dS black hole, no Casimir energy exists, for
odd--dimensional cases, the Casimir energy depends upon the topology and
geometry of spacetime foliations of the bulk near the conformal boundary %
\cite{IDA}. Insertion of (\ref{massesonn}) yields this value explicitly for
TB$^{\pm }$. Figure \ref{fig1} illustrates the behavior of $\mathfrak{M}%
^{\pm }/\ell $ at future infinity, as a function of $n/\ell $ for $q=1$ and $%
3$.

While TB$^{-}$ has $\mathfrak{M}^{-}<0$ for positive $n>n_{c}$ ($n_{c}$
depends on $q$ and by increasing $q,$ it reduces to zero), TB$^{+}$ has $%
\mathfrak{M}^{+}>\mathfrak{M}^{\text{dS}}=0$ and so forms a class of
spacetimes that are counterexamples to the Balasubramanian \textit{et al.}
conjecture \cite{bala}, since $\mathfrak{M}^{+}>0$\ and there are no
cosmological singularities. The signs of $\mathfrak{M}^{\pm }$ are reversed
for negative $n$, in which case TB$^{-}$ violates the conjecture. 

The total action (\ref{totaction}) of the TB$^{\pm }$ spacetimes is 
\begin{equation}
I^{\pm }=-\frac{\beta _{H}}{2\ell ^{2}}(m^{\pm }\ell ^{2}+\left( \tau ^{\pm
}\right) ^{3}+3n^{2}\tau ^{\pm })  \label{acttot}
\end{equation}%
where $\beta _{H}=\left| \frac{-4\pi }{V^{\prime }(\tau )}\right| _{\tau
=\tau _{0}}=$ $\left| \frac{8\pi n}{q}\right| $ is the analogue of the
Hawking temperature outside of the cosmological horizon. The parameters $%
m^{\pm }$ and $\tau ^{\pm }$ in (\ref{acttot}) are given by (\ref{massesonn}%
) and (\ref{taus}).

In the present case, from the relation (\ref{entr}), we obtain 
\begin{equation}
S^{\pm }=-\frac{\beta _{H}}{2\ell ^{2}}(m^{\pm }\ell ^{2}-\left( \tau ^{\pm
}\right) ^{3}-3n^{2}\tau ^{\pm })  \label{S}
\end{equation}%
and it is straightforward to show that the first law $dS^{\pm }=\beta _{H}d%
\mathfrak{M}^{\pm }$ is satisfied for TB$^{\pm }$ respectively.\ Figure \ref%
{fig2} shows the behavior of $\mathcal{S}^{\pm }=S^{\pm }/\ell ^{2}$ as a
function of $n/\ell $ for $q=1,3$.

For $q=1$ the entropy $\mathcal{S}^{-}\left( n\right) =-\mathcal{S}%
^{-}\left( -n\right) $ attains for increasing $n$ a positive maximum near $%
n=0.116$, vanishes near $n=0.175$, attains a negative local minimum at $%
n=0.506$, increasing again to positive values for $n>0.590$ before attaining
a global positive maximum at the critical Bolt charge $n_{c}=\frac{1}{6}%
\sqrt{6+3\sqrt{5}}\approx 0.594$. For all values of Bolt charge the $\mathbf{%
N}$--bound \cite{bousso} on the entropy ($\mathcal{S}^{-}\leq \pi $ ) is
satisfied. For the other values of $q$, the $\mathbf{N}$--bound on the
entropy also is satisfied. So for TB$^{-}$ and $q=1$, the $\mathbf{N}$%
--bound is satisfied everywhere while the mass conjecture holds only for
positive Bolt charge $n>n_{c}$.

Similar considerations for TB$^{+}$ imply that for positive Bolt charge $n$
and any value of $q,$ both the $\mathbf{N}$--bound and the mass conjecture
are violated. Indeed, the quasi-regular singularity structure of the
spacetime is unaltered for any choice of these parameters. 
\begin{figure}[tbp]
\begin{center}
\begin{minipage}[c]{.45\textwidth}
         \centering
         \includegraphics[width=\textwidth]{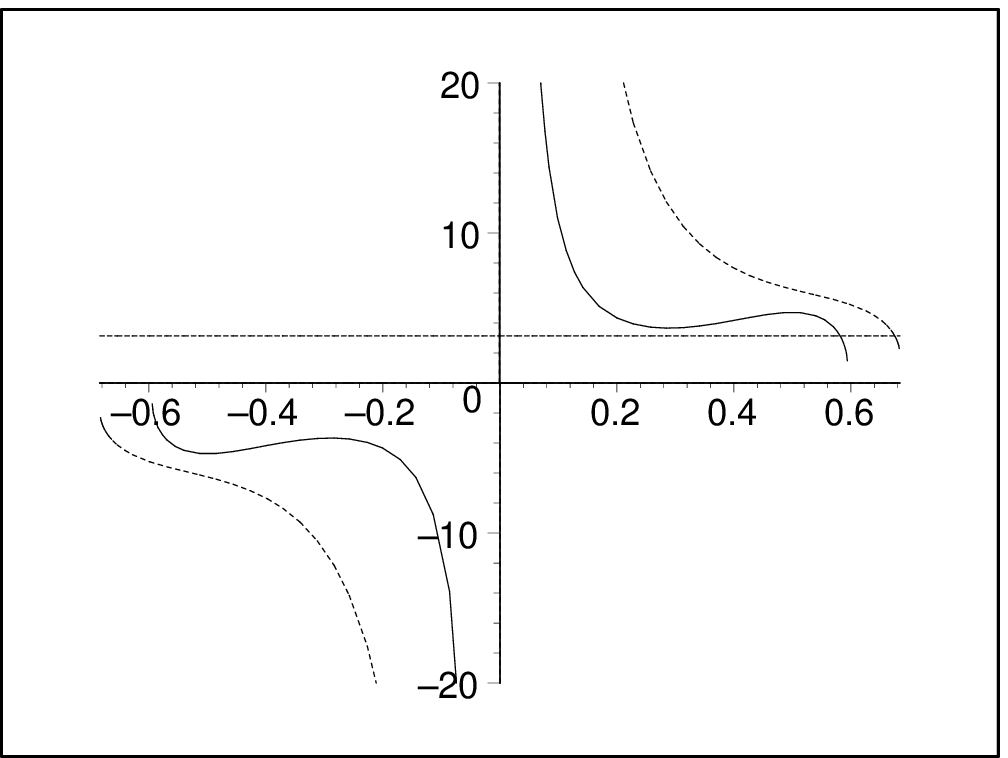}
         \label{SPLUSK13BW}
     \end{minipage}%
\begin{minipage}[c]{.45\textwidth}
         \centering
         \includegraphics[width=\textwidth]{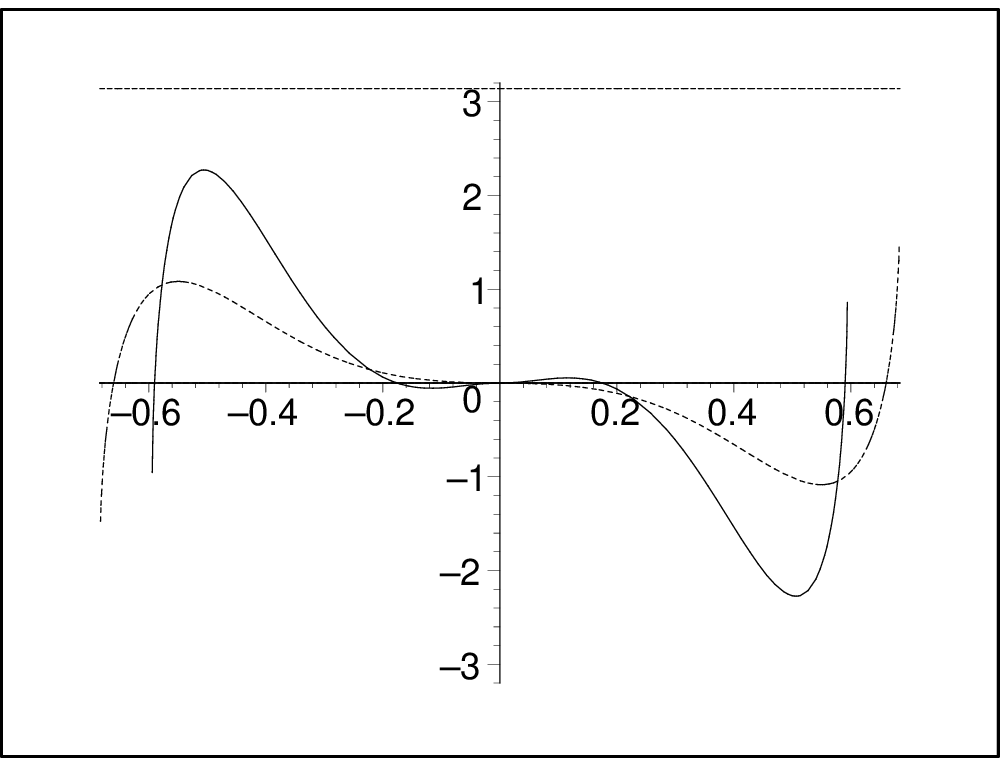}
         \label{SMINUSK13BW}
     \end{minipage}
\end{center}
\caption{Left/right : Entropy $\mathcal{S}^{+/-}$ of TB$^{+/-}$ versus $%
n/\ell $ with $q=1$ (solid) and $q=3$ (dotted). The horizontal dashed lines
denote the $\mathbf{N}$--bound.}
\label{fig2}
\end{figure}

\begin{figure}[tbp]
\begin{center}
\begin{minipage}[c]{.45\textwidth}
         \centering
         \includegraphics[width=\textwidth]{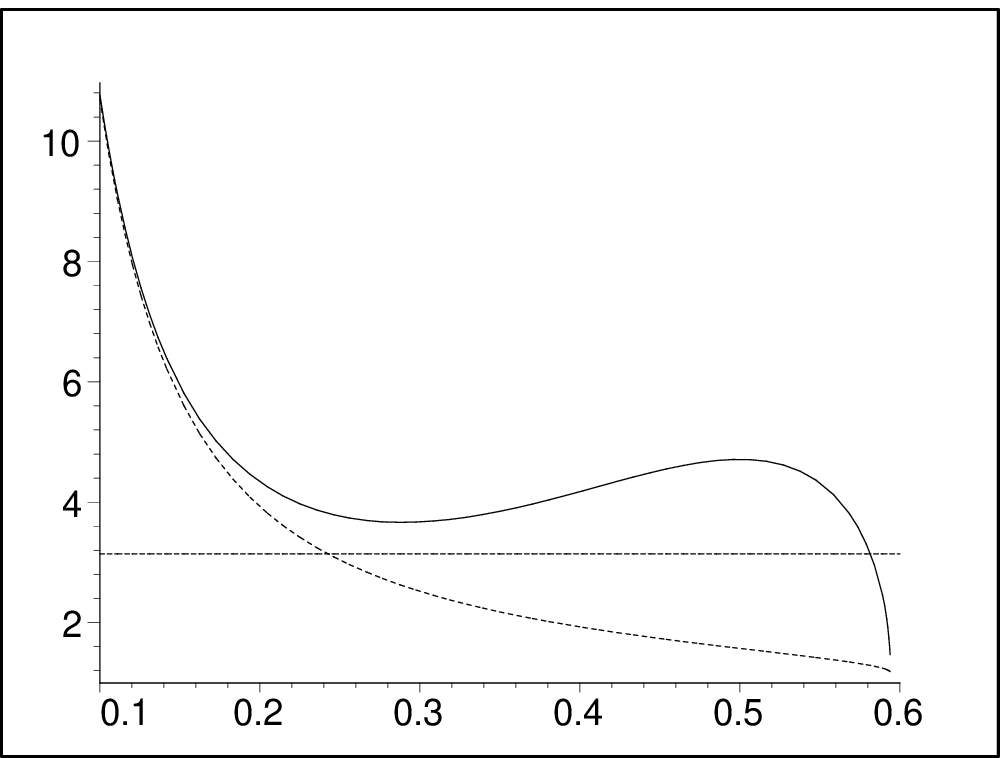}
         \label{Entropiesp}
     \end{minipage}%
\begin{minipage}[c]{.45\textwidth}
         \centering
         \includegraphics[width=\textwidth]{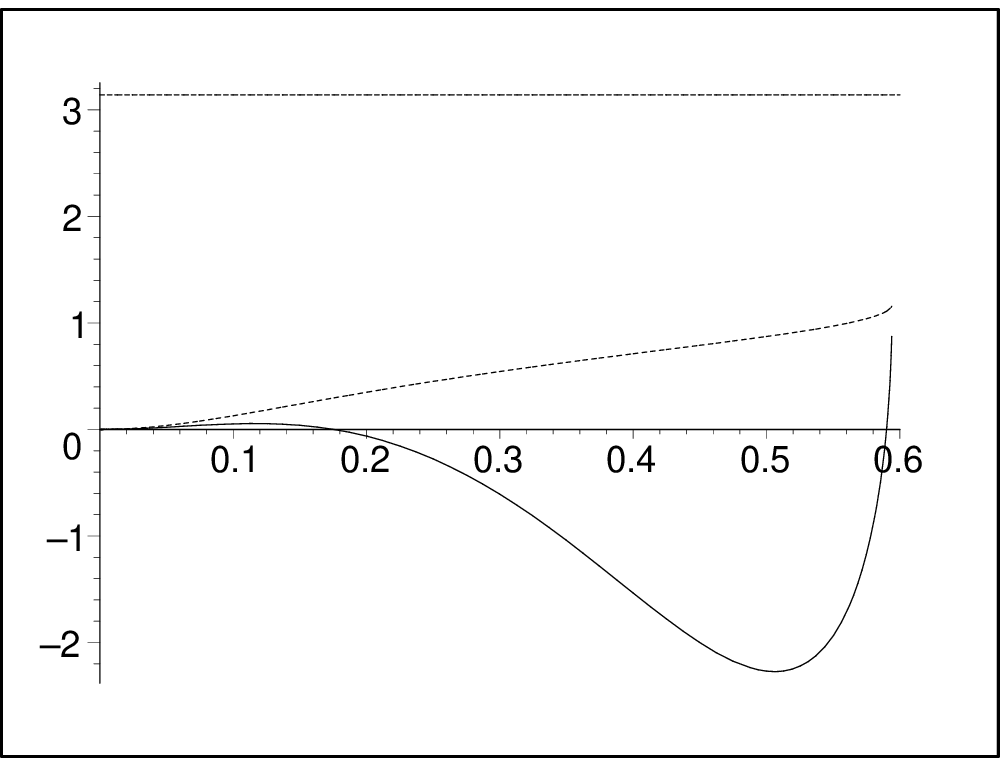}
         \label{Entropiesm}
     \end{minipage}
\end{center}
\caption{Left/right : Gibbs-Duhem entropy $\mathcal{S}^{+/-}$ (solid),
cosmological entropy (dotted) and $\mathbf{N}$-bound (dashed) for positive
NUT charge of TB$^{+/-}$.}
\label{fig22}
\end{figure}

Since the usual relationship between entropy and area does not hold for
NUT--charged spacetimes \cite{misner,NUTrefs} it is natural to inquire if
the horizon area of dS spacetime is maximal. We find that while the (fixed--$%
t$) area of the cosmological horizon of TB$^{-}$ is always less than that of
pure dS spacetime, that of TB$^{+}$ exceeds it for $n<0.2425$. Consequently
if one interprets the $\mathbf{N}$--bound in terms of a relationship between
horizon areas (as opposed to entropies), we still find that (within this
range) the $\mathbf{N}$--bound is violated. For TB$^{-}$,the fixed--$t$ area
of the cosmological horizon is less than the cosmological horizon area of
pure dS spacetime for all values of NUT charge, and so the re-interpreted $%
\mathbf{N}$--bound is respected. In the TB$^{+}$case, the Gibbs--Duhem
entropy is larger than one--quarter of the horizon area, which in turn is
larger than $\pi \ell ^{2}$, the entropy of pure dS spacetime. In the TB$%
^{-} $case, these inequalities are reversed, with $\pi \ell ^{2}$always
greater than the Gibbs--Duhem entropy, and both the Gibbs--Duhem entropy and
one--quarter the area of cosmological horizon respect the $\mathbf{N}$%
--bound. Figure \ref{fig22} shows the behaviour of entropies for the TB$^{+}$%
and TB$^{-}$ cases. 

It is worth mentioning briefly the results of the conserved quantities for
four--dimensional dS inflationary and covering patches and the
Schwarzschild--dS black hole. The action of the dS spacetimes in
inflationary coordinates (in both big-bang and big-crunch patches) is
finite, but in covering coordinates a linear divergence remains in odd
dimensions that cannot be cancelled by local terms that are polynomial in
boundary curvature invariants. As we mentioned after equation (\ref{totmass}%
), the conserved mass of the four--dimensional Schwarzschild--dS black hole
is given by the negative of the black hole mass parameter and so the
conserved mass is always negative and respects the maximal mass conjecture.
It has been shown in general \cite{GM} that the conserved mass and action of
the Schwarzschild--dS black holes with different dimensionalities (up to
nine dimensions) are finite. Moreover the entropy of the Schwarzschild--dS
black hole in four dimensions respects the $\mathbf{N}$--bound.

\begin{figure}[tbp]
\centering       
\begin{minipage}[c]{.45\textwidth}
         \centering
         \includegraphics[width=\textwidth]{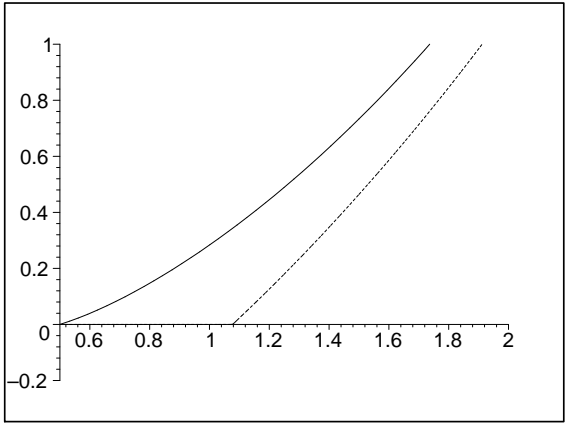}
         \caption{$c$--function of TB$^+$ solution versus
$\tau \geq \tau ^{+}=0.50$ for $q=1$ (solid) and $\tau \geq \tau ^{+}=1.077$
for $q=3$ (dotted). }
         \label{CPLUS}
\end{minipage}
\begin{minipage}[c]{0.2\textwidth}
\end{minipage}
\begin{minipage}[c]{.45\textwidth}
         \centering
         \includegraphics[width=\textwidth]{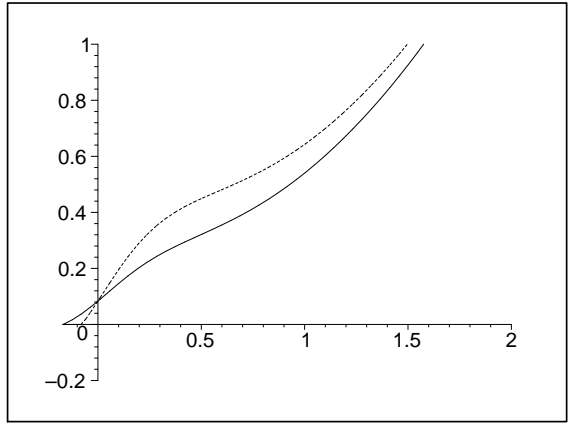}
         \caption{$c$--function of TB$^-$ solution versus
$\tau \geq \tau ^{-}=-0.167$ for $q=1$ (solid) and $\tau \geq \tau ^{-}=-0.077$
for $q=3$ (dotted).}
         \label{CMINUS}
\end{minipage}\label{fig55}
\end{figure}

In both the AdS and dS cases there is a natural correspondence between
phenomena occurring near the boundary (or in the deep interior) of either
spacetime and UV (IR) physics in the dual CFT. Solutions that are
asymptotically (locally) dS lead to an interpretation in terms of
renormalization group flows and an associated generalized dS $c$--theorem.
This theorem states that in a contracting patch of dS spacetime, the
renormalization group flows toward the infrared and in an expanding
spacetime, it flows toward the ultraviolet. Since the spacetime (\ref{TBDS})
is asymptotically (locally) dS, we can use the four-dimensional $c$%
--function \cite{Leb} 
\begin{equation}
c=\left( G_{\mu \nu }n^{\mu }n^{\nu }\right) ^{-1}=\frac{1}{G_{\tau \tau }}
\label{cfunction}
\end{equation}%
where $n^{\mu }$ is the unit normal vector to a constant $\tau $--slice. In
figs. \ref{CPLUS} and \ref{CMINUS}, the diagrams of the TB$^{\pm }$
spacetime $c$--functions outside the cosmological horizon with $\ell =1$ and 
$n=0.5$ for two cases $q=1$ and $3$ are plotted. 

As one can see from these figures, outside the cosmological horizon, the $c$%
--function is a monotonically increasing function of $\tau $, indicative of
the expansion of a constant $\tau $--surface of the metric (\ref{TBDS})
outside of the cosmological horizon. Since the metric (\ref{TBDSgen}) at
future infinity $\tau \rightarrow +\infty $, reduces to 
\begin{equation}
ds_{R}^{2}\rightarrow -du^{2}+e^{2u/\ell }d\Sigma _{3}^{2}  \label{TNDSbou}
\end{equation}%
where $u=\ell \ln \tau $ and $d\Sigma _{3}^{2}$ is the metric of
three-dimensional constant $u$--surface, the scale factor in (\ref{TNDSbou})
expands exponentially near future infinity. Hence the behavior of the $c$%
--function in figures (\ref{CPLUS}) and (\ref{CMINUS}) is in good agreement
with what one expects from the $c$--theorem. According to the $c$--theorem,
for any asymptotically (locally) dS spacetimes, the $c$--function must
increase (decrease) for any expanding (contracting) patch of the spacetime.

\subsection{Six dimensional case}

\begin{figure}[tbp]
\begin{center}
\begin{minipage}[c]{.45\textwidth}
         \centering
         \includegraphics[width=\textwidth]{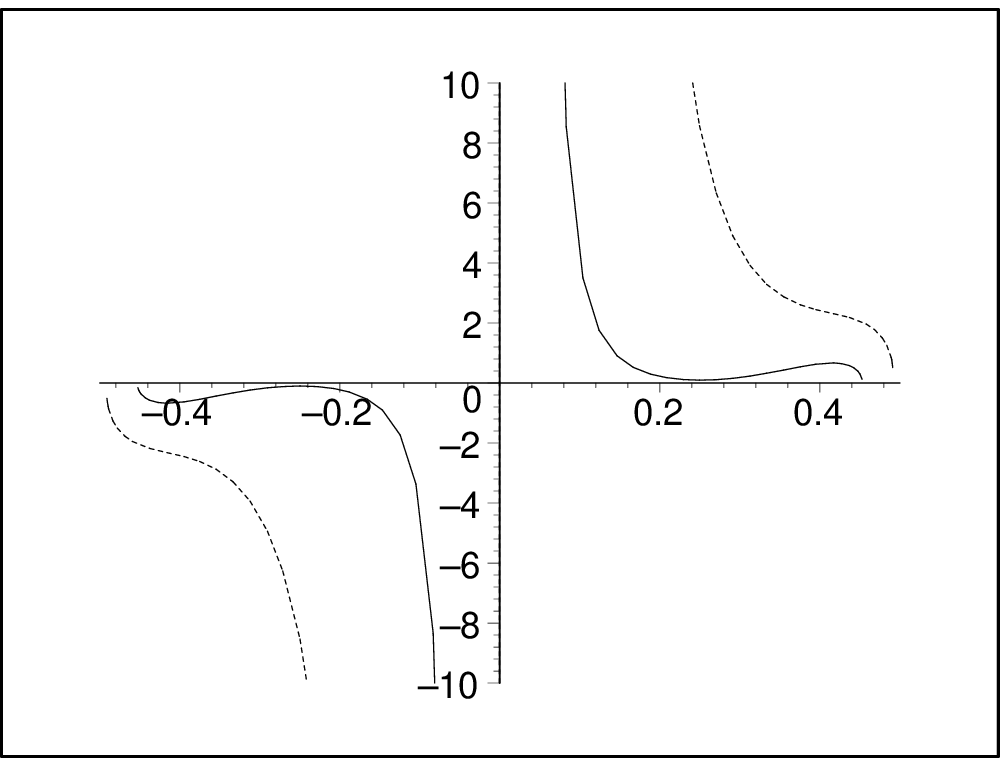}
\end{minipage}
\begin{minipage}[c]{.45\textwidth}
         \centering
         \includegraphics[width=\textwidth]{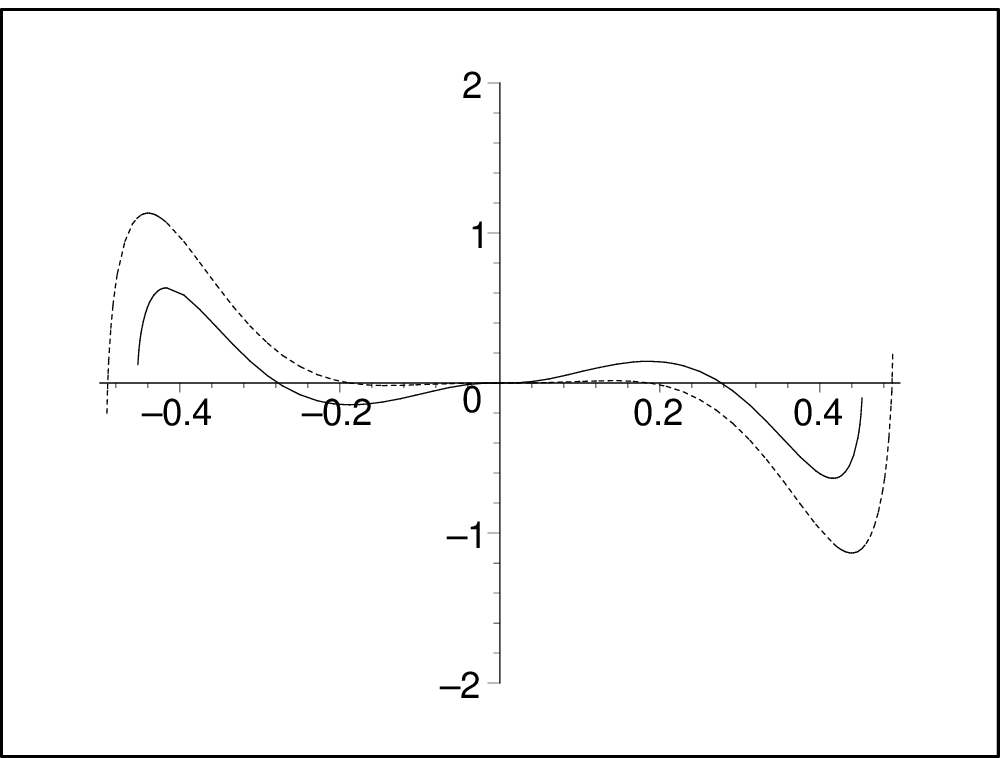}
\end{minipage}
\end{center}
\caption{Left/right: Mass of TB$_{6}^{+/-}$ with $q=1$ (solid) and $q=3$
(dotted).}
\label{mass6pos}
\end{figure}

\begin{figure}[tbp]
\begin{center}
\begin{minipage}[c]{.45\textwidth}
         \centering
         \includegraphics[width=\textwidth]{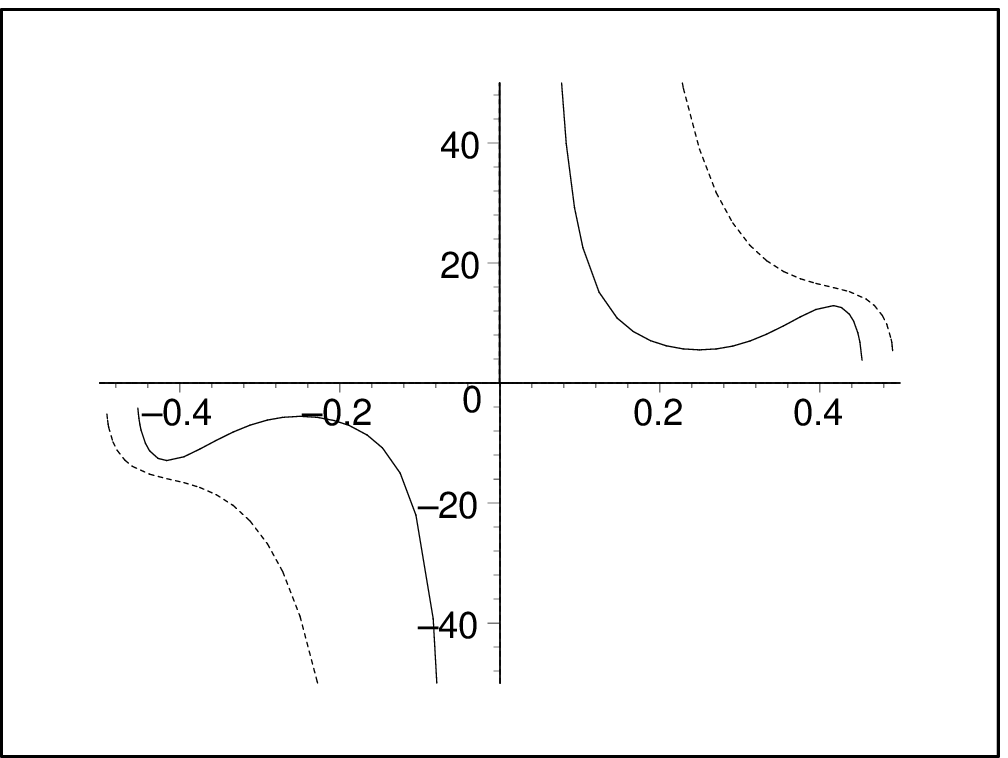}
\end{minipage}
\begin{minipage}[c]{.45\textwidth}
         \centering
         \includegraphics[width=\textwidth]{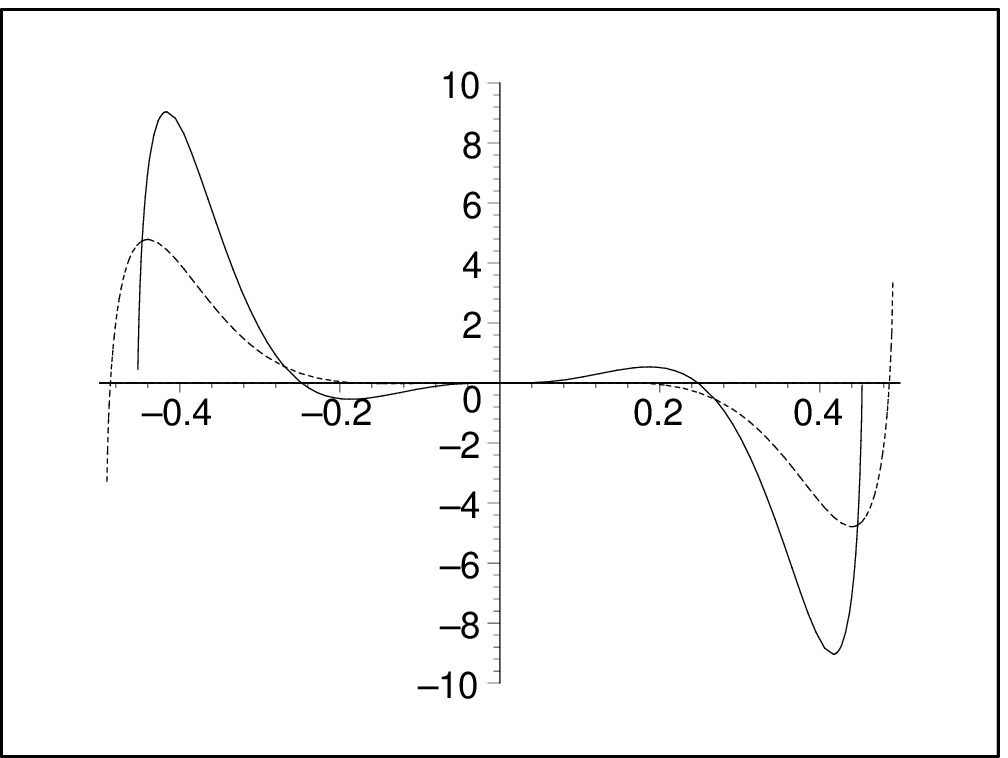}
\end{minipage}
\end{center}
\caption{Left/right: Entropy of TB$_{6}^{+/-}$ with $q=1$ (solid) and $q=3$
(dotted).}
\label{ent6pos}
\end{figure}

The ($5+1$) dimensional Taub--Bolt--dS metric, outside the cosmological
horizon, is given by the line element 
\begin{eqnarray}
ds_{6}^{2} &=&V(\tau )\left[ dt+2n\left( \cos (\theta _{1})d\phi
_{1}^{2}+\cos (\theta _{2})d\phi _{2}^{2}\right) \right] ^{2}-\frac{d\tau
^{2}}{V(\tau )}  \notag \\
&&+(\tau ^{2}+n^{2})\left( d\theta _{1}^{2}+\sin ^{2}(\theta _{1})d\phi
_{1}^{2}+d\theta _{2}^{2}+\sin ^{2}(\theta _{2})d\phi _{2}^{2}\right) .
\label{TBDSgen}
\end{eqnarray}%
The metric function $V(\tau )$ is 
\begin{equation}
V(\tau )=\frac{3\tau ^{6}+(-\ell ^{2}+15n^{2})\tau ^{4}+3n^{2}(-2\ell
^{2}+15n^{2})\tau ^{2}-3n^{4}(-\ell ^{2}+5n^{2})+6m\tau \ell ^{2}}{3(\tau
^{2}+n^{2})^{2}\ell ^{2}}  \label{FtBolt6}
\end{equation}%
where $n$ is the non-vanishing NUT charge and $\Lambda ={\frac{10}{\ell ^{2}}%
}$. The coordinate $t$ parameterizes an $S^{1}$ Hopf fibration over the
non--vanishing $S^{2}\otimes S^{2}$ base space, parameterized by $(\theta
_{1},\phi _{1},\theta _{2},\phi _{2})$. It must have periodicity$\frac{12\pi
\left| n\right| }{q}$ to avoid conical singularities, where $q$ is a
positive integer. The geometry of a constant--$\tau $ surface is that of a
Hopf fibration of $S^{1}$ over $S^{2}\otimes S^{2}$ which is a well defined
hypersurface in spacetime where $V(\tau )>0$ is outside of the past/future
cosmological horizons. The spacelike Killing vector $\partial /\partial t$
has a fixed point set where $V(\tau _{c})=0$ whose topology is that of an $%
S^{2}\otimes S^{2}$ base space. Since $\frac{\partial }{\partial \phi _{1}}$
and $\frac{\partial }{\partial \phi _{2}}$ are Killing vectors, for any
constant $(\phi _{1},\phi _{2})$--slice near the horizon $\tau =\tau _{c}$
additional conical singularities will be introduced in the $(t,\tau )$
Euclidean section unless $t$ has the period 
\begin{equation}
\beta _{6d}=\frac{4\pi }{\left| V^{\prime }(\tau _{c})\right| }.
\label{betatb6}
\end{equation}%
This period must be equal to $\frac{12\pi \left| n\right| }{q}$, which
forces $\tau _{c}=\tau _{c}^{\pm }$\ where 
\begin{equation}
\tau _{c}^{\pm }=\frac{q\ell ^{2}\pm \sqrt{q^{2}\ell
^{4}-900n^{4}+180n^{2}\ell ^{2}}}{30n}  \label{tau6}
\end{equation}%
and we denote the respective extensions by TN$_{6}^{\pm }$. We note the
spacetime exists only for the following NUT charges: 
\begin{equation}
\left| n\right| \leq \ell \frac{\sqrt{90+30\sqrt{q^{2}+9}}}{30}.
\label{nut6}
\end{equation}%
\bigskip The mass parameters are given by 
\begin{equation}
m=-\frac{3\tau _{c}^{6}-\tau _{c}^{4}(\ell ^{2}+15n^{2})-\tau
_{c}^{2}n^{2}(6\ell ^{2}-45n^{2})+3n^{4}(\ell ^{2}-5n^{2})}{6\ell ^{2}\tau
_{c}}.  \label{masesTB6}
\end{equation}%
The conserved mass and the action near future infinity are found to be \cite%
{CGM2} 
\begin{equation}
\mathfrak{M}_{6d}=-8\pi m_{R}-\frac{\pi }{54\ell ^{2}\tau }(2205n^{4}\ell
^{2}-10773n^{6}-\ell ^{6}-63n^{2}\ell ^{4})+\mathcal{O}\left( \frac{1}{\tau
^{2}}\right)  \label{TBmass6}
\end{equation}

\begin{equation}
I_{6d}=-\frac{2\beta _{6d}\pi }{3\ell ^{2}}(3\tau _{c}^{5}+10n^{2}\tau
_{c}^{3}+15n^{4}\tau _{c}+3m_{R}\ell ^{2})+\mathcal{O}\left( \frac{1}{\tau }%
\right) .  \label{TBaction6}
\end{equation}%
Applying the Gibbs--Duhem relation (\ref{entr}), the total entropy at future
infinity is 
\begin{equation}
S_{6d}=\frac{2\pi \beta _{R,6d}(3\tau _{c}^{5}+10n^{2}\tau
_{c}^{3}+15n^{4}\tau _{c}-9m_{R}\ell ^{2})}{3\ell ^{2}}  \label{TBentropy6}
\end{equation}%
where $\beta_{6d}$ is given by: 
\begin{equation}
\beta _{6d}=\frac{6\pi (\tau _{c}^{2}+n^{2})^{3}\ell ^{2}}{\left| 3\tau
_{c}^{7}+9\tau _{c}^{5}n^{2}+\tau _{c}^{3}n^{2}(4\ell ^{2}-15n^{2})-9m\ell
^{2}\tau _{c}^{2}+n^{4}\tau _{c}(75n^{2}-12\ell ^{2})+3m\ell
^{2}n^{2}\right| }.  \label{beta6}
\end{equation}%
Figures \ref{mass6pos} and \ref{ent6pos} show the conserved masses and
entropies for two different branches of six dimensional TB$_{6}^{\pm }$
spacetimes with $q=1$ and $q=3$.

Figure \ref{mass6pos} shows that for all positive NUT charge, TN$_{6}^{+}$
has a positive mass. The mass of TN$_{6}^{-}$ (with $q=1$) for $%
n<0.27731503405\ell $ is also positive. 
Using equations (\ref{TBmass6}), (\ref{masesTB6}) and (\ref{tau6}), the
conserved mass is 
\begin{eqnarray}
\mathfrak{M}_{6d}^{\pm }(\tau _{0}=\tau _{0}^{\pm }) &=&\frac{\pm 2\pi }{%
759375n^{5}\ell ^{2}}\{\sqrt{q^{2}\ell ^{4}+180n^{2}\ell ^{2}-900n^{4}}%
(810000n^{8}-54000n^{6}\ell ^{2}  \notag \\
&&-1350n^{4}\ell ^{4}+450n^{4}q^{2}\ell ^{4}+60n^{2}q^{2}\ell ^{6}+q^{4}\ell
^{8})  \label{TBDS6mass} \\
&&\pm 150n^{2}q^{3}\ell ^{8}\pm q^{5}\ell ^{10}\}  \notag
\end{eqnarray}%
and from (\ref{TBentropy6}), the entropy is 
\begin{eqnarray}
S_{6d}^{\pm } &=&\frac{\pm \pi ^{2}(\pm q^{2}\ell ^{2}\pm 90n^{2}+q\sqrt{%
q^{2}\ell ^{4}+180n^{2}\ell ^{2}-900n^{4}})^{3}}{101250n^{4}q\ell ^{2}}%
\{(q^{4}\ell ^{8}  \notag \\
&&+90n^{2}\ell ^{6}q^{2}+300n^{4}q^{2}\ell ^{4}-27000n^{6}\ell
^{2}+540000n^{8})  \notag \\
&&\sqrt{q^{2}\ell ^{4}+180n^{2}\ell ^{2}-900n^{4}}\mp 150n^{4}q^{3}\ell
^{6}\pm 4050n^{4}\ell ^{6}q\pm q^{5}\ell ^{10}\pm 180n^{2}\ell ^{8}q^{3}\}/ 
\notag \\
&&\{\pm (60750n^{6}q^{2}-270n^{2}q^{4}\ell ^{4}+675n^{4}q^{4}\ell
^{2}-q^{6}\ell ^{6}-182250n^{6})-  \notag \\
&&\sqrt{q^{2}\ell ^{4}+180n^{2}\ell ^{2}-900n^{4}}(q^{5}\ell
^{4}+6075n^{4}q+180n^{2}\ell ^{2}q^{3}-225n^{4}q^{3})\}.
\label{TBDS6entropy}
\end{eqnarray}%
The entropy for both branches satisfies the first law $dS_{6d}^{\pm }=\beta
_{6d}^{\pm }d\mathfrak{M}_{6d}^{\pm }$.

\begin{figure}[tbp]
\begin{center}
\begin{minipage}[c]{.45\textwidth}
         \centering
         \includegraphics[width=\textwidth]{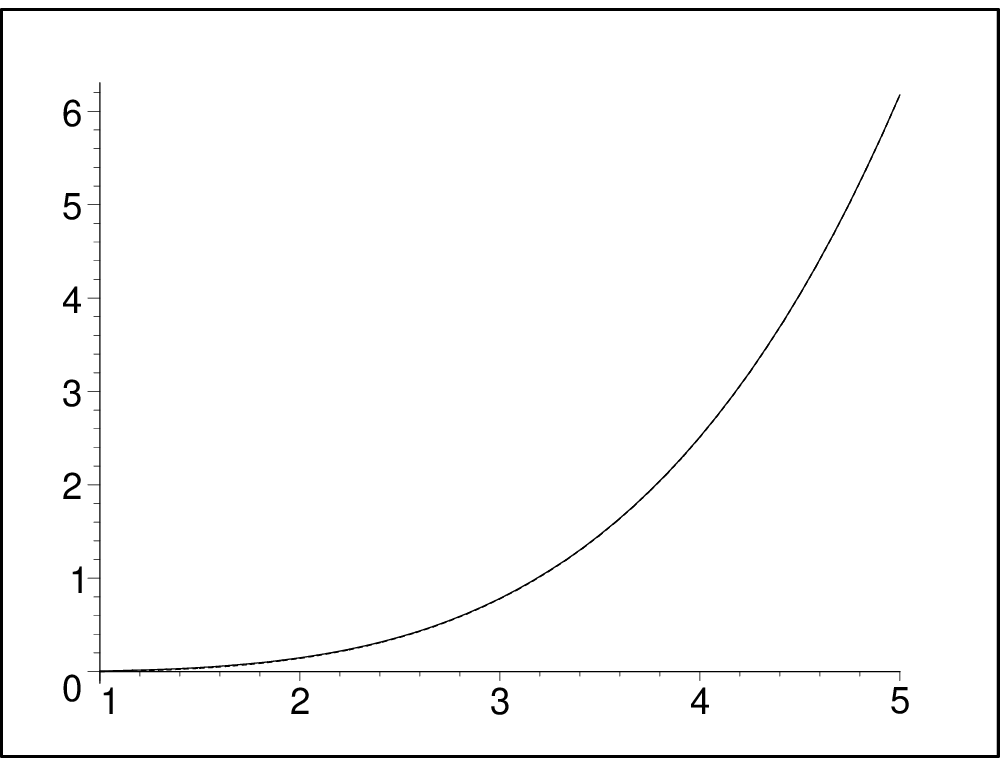}
\end{minipage}
\begin{minipage}[c]{.45\textwidth}
         \centering
         \includegraphics[width=\textwidth]{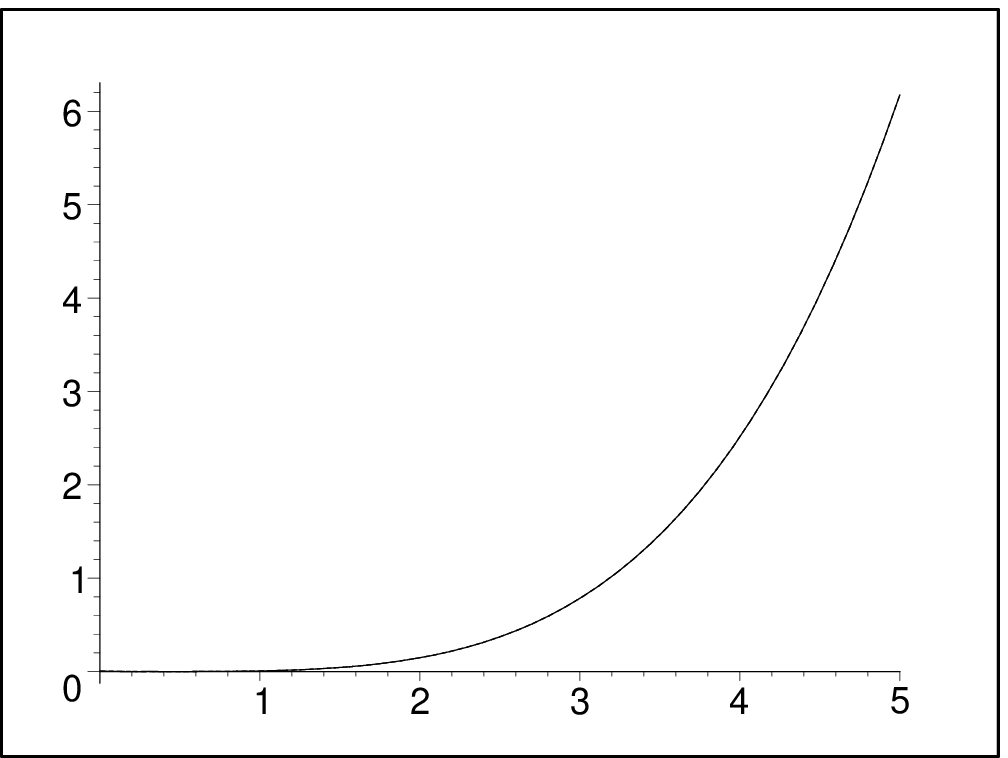}
\end{minipage}
\end{center}
\caption{Left/right: $c$--function of TB$_{6}^{+/-}$ solution versus $%
\protect\tau $ with different values of $q=1$ (solid) and $q=3$ (dotted).
The two plots overlap in most parts of $\protect\tau $--axis.}
\label{C6pos}
\end{figure}

The six-dimensional $c$--function is given by 
\begin{equation}
c=\left( G_{\mu \nu }n^{\mu }n^{\nu }\right) ^{-2}=\frac{1}{(G_{\tau \tau
})^{2}}  \label{cfunction6d}
\end{equation}%
where $n^{\mu }$ is the unit normal vector to a constant $\tau $--slice. In
figure \ref{C6pos}, the diagrams of the TB$_{6}^{\pm }$ spacetimes $c$
-functions outside the cosmological horizon with $\ell =1$ and $n=0.25$ for
two cases $q=1$ and $3$ are plotted. 

\begin{figure}[tbp]
\begin{center}
\begin{minipage}[c]{.43\textwidth}
         \centering
         \includegraphics[width=\textwidth]{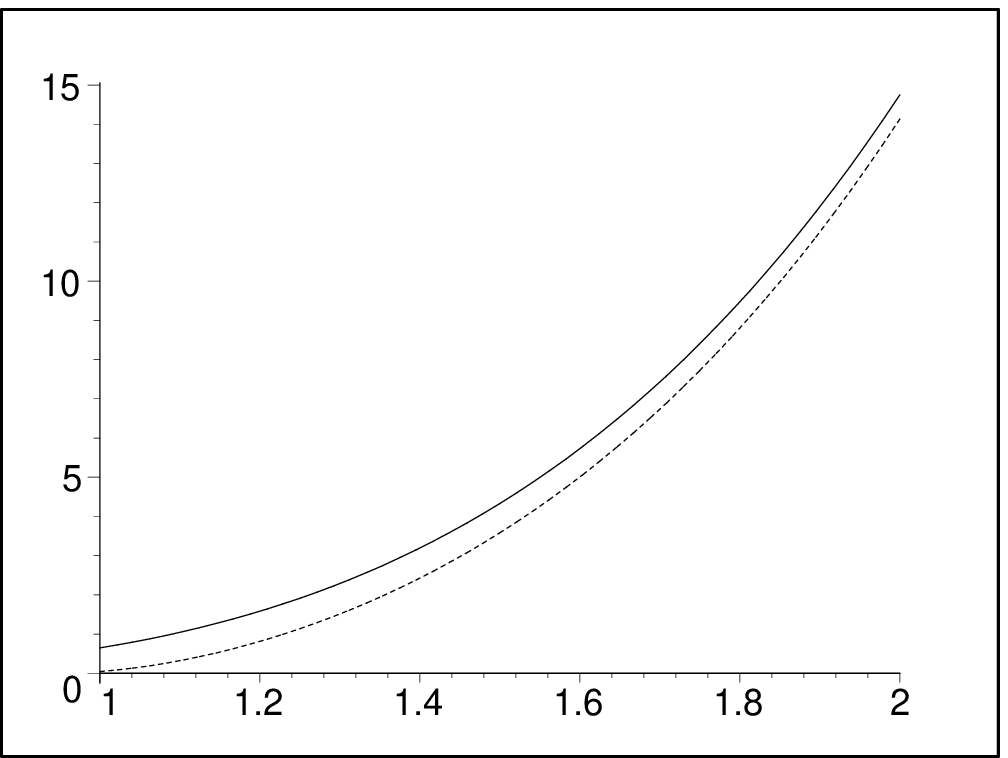}
\end{minipage}%
\begin{minipage}[c]{.43\textwidth}
         \centering
         \includegraphics[width=\textwidth]{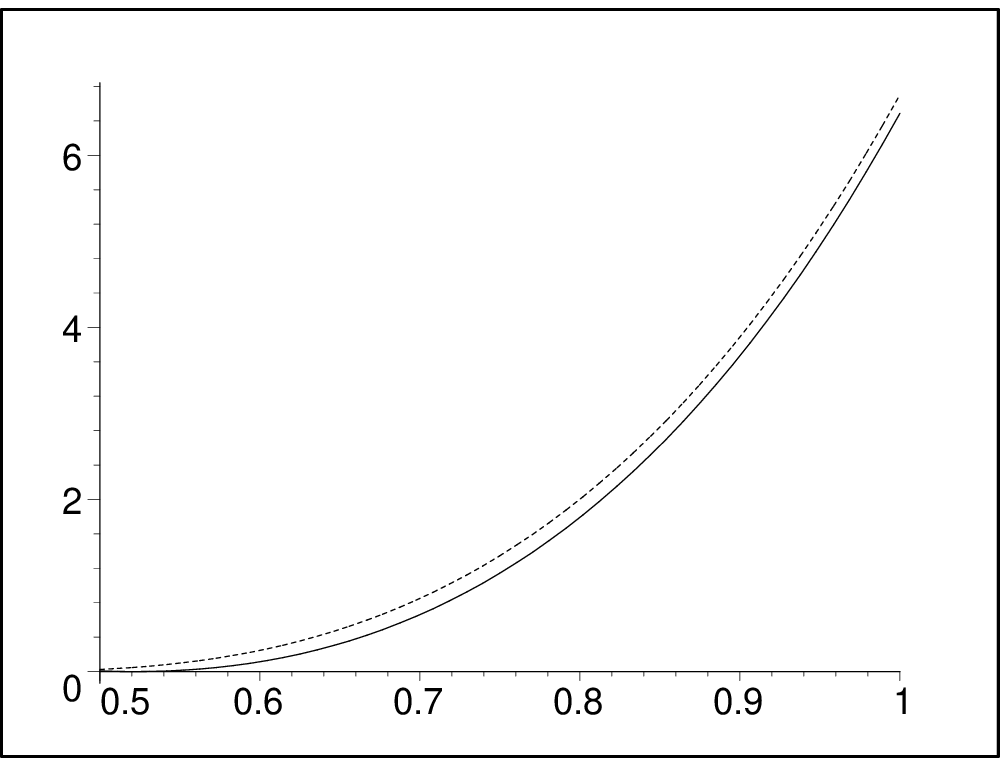}
\end{minipage}
\end{center}
\caption{Left/right: Fine structure of $c$--function of TB$_{6}^{+/-}$
solution versus $\protect\tau $ with different values of $q=1$ (solid) and $%
q=3$ (dotted). Note that the values on the vertical axis for the left
diagram are in units of $10^{-2}$ and for the right diagram are in units of $%
10^{-3}$.}
\label{C6posfine}
\end{figure}

As one can see from these figures, outside the cosmological horizon, the $c$%
--function is a monotonically increasing function of coordinate $\tau $,
showing the expansion of a constant $\tau $--surface of the metric (\ref%
{TBDSgen}) outside of the cosmological horizon. We note that the behavior of
the $c$--function is rather insensitive to $q$. Figure \ref{C6posfine} shows
finer diagrams of $c$--functions for TB$_{6}^{\pm }$ spacetimes.

\section{Examples --- C approach}

\label{sec:examplesC}

We also wish to present the calculation for the C--approach, where we
Wick--rotate the time and NUT charge ($t\rightarrow \text{i}T,n\rightarrow 
\text{i}N$) in analogy with the AdS case \cite{CFM}. Here, also, there are
regions where the $\mathbf{N}$--bound and maximal--mass conjecture are
violated for certain values of the NUT charge in four dimensions. Note that
unlike the R--approach method, this method produces a NUT and a Bolt
solution similar to the Taub--NUT--AdS case \cite{CFM,AC}. Also, note that
the discussions above concerning the general structure and singularities of
the spacetime apply here also, and so we won't repeat them.

\subsection{Four Dimensional Case}

The metric in this case takes the following form 
\begin{equation}
ds_{C}^{2}=-F(\rho )\left( dT+2N\cos (\theta )d\phi \right) ^{2}-\frac{d\rho
^{2}}{F(\rho )}+(\rho ^{2}-N^{2})\left( d\theta ^{2}+\sin ^{2}(\theta )d\phi
^{2}\right)  \label{mtrc4dC}
\end{equation}%
where 
\begin{equation}
F(\rho )=\frac{\rho ^{4}-(\ell ^{2}+6N^{2})\rho ^{2}+2m\rho \ell
^{2}-N^{2}(\ell ^{2}+3N^{2})}{(\rho ^{2}-N^{2})\ell ^{2}}.  \label{F4dC}
\end{equation}%
$N$ is the non--vanishing NUT charge, and the cosmological constant is given
by $\Lambda ={\textstyle\frac{3}{\ell ^{2}}}$. Note, though, that the
signature of the metric in this case is $(--++)$, and so the geometry is no
longer strictly a Hopf fibration of $S^{1}$ over a 2--sphere, since the
coordinate $T$ is now timelike. This brings into question the physical
relevance of this form of the metric. However, the metric is independent of
the $T$ coordinate, and so we can still calculate the action and
thermodynamic quantities. We will do so, while keeping in mind the above.

$T$ parameterizes a circle fibred over the non--vanishing sphere
parameterized by $(\theta,\phi)$, and thus this case must also have a
periodicity condition 
\begin{equation}
\beta_{4C} = \frac{4\pi}{|F^{\prime}(\rho)|} = \frac{8\pi |N|}{q}
\label{beta4C}
\end{equation}
imposed in order to avoid conical singularities, where $q$ is again a
positive integer.

Using the method of counter--terms for de Sitter space \cite{GM}, one can
find the action in four dimensions (before specifying a NUT or Bolt
solution) 
\begin{equation}
I_{C,4d}=\frac{\beta _{4C}\left( \rho _{+}^{3}-3N^{2}\rho _{+}+m\ell
^{2}\right) }{2\ell ^{2}}  \label{I4C}
\end{equation}%
with $\rho _{+}$ the largest positive root of $F(\rho )$, determined by the
fixed point set of $\partial _{T}$. The mass parameter $m$ will have
different values for different $\rho \geq \rho _{+}$.

Working at future infinity, using (\ref{Mcons}) with the full $T_{ab}$,
including counter-terms, the conserved mass for the C--approach is given by 
\begin{equation}
\mathfrak{M}_{C,4d}=-m+\frac{105N^{4}+30N^{2}\ell ^{2}+\ell ^{4}}{8\ell
^{2}\rho }+\mathcal{O}\left( \frac{1}{\rho ^{2}}\right) .  \label{mcons4dC}
\end{equation}%
Note that this reduces to the pure Schwarzschild--dS result \cite{GM} for $%
N=0$. Finally, by applying the Gibbs-Duhem relation (\ref{GDoutfinal}) the
total entropy can be calculated: 
\begin{equation}
S_{C,4d}=\frac{\beta _{4C}\left( \rho _{+}^{3}-3N^{2}\rho _{+}-m\ell
^{2}\right) }{2\ell ^{2}}.  \label{S4C}
\end{equation}%
This entropy satisfies the first law of gravitational thermodynamics for
both the NUT and Bolt cases (see below).

These formula are generic, and the metric and these equations give two
solutions, depending on the fixed point set of $\partial_T$. These arise
from (\ref{beta4C}). If $\rho_+ = N$, $F(\rho=N) = 0$ and the fixed point
set is zero-dimensional, giving the ``NUT''; if $\rho_+ = \rho_{b\pm} > N$,
the fixed point set is two-dimensional, giving the ``Bolt''. Each case gives
different results, both of which are of interest.

\subsubsection{NUT solution}

The NUT mass parameter can be solved for from (\ref{beta4C}) to give 
\begin{equation}
m_{4C,n}=\frac{N(\ell ^{2}+4N^{2})}{\ell ^{2}}.  \label{mnut4C}
\end{equation}%
This mass is always positive, and so the conserved mass $\mathfrak{M}=-m$
for the NUT solution will always be less than pure de Sitter, satisfying the
maximal mass conjecture for all values of $N$. The flat space limit ($\ell
\rightarrow \infty $) gives $m_{4C,n}\rightarrow N$, and the high
temperature limit $N\rightarrow 0$ gives $m_{4C,n}\rightarrow 0$.

The period in four dimensions, with $q=1$, is $\beta _{4C,n}=8\pi N$, giving
the NUT action and entropy 
\begin{eqnarray}
I_{4C,NUT} &=&-\frac{4\pi N^{2}(\ell ^{2}+2N^{2})}{\ell ^{2}}  \label{Inut4C}
\\
S_{4C,NUT} &=&-\frac{4\pi N^{2}(\ell ^{2}+6N^{2})}{\ell ^{2}}.
\label{Snut4C}
\end{eqnarray}%
It can be shown that (\ref{Snut4C},\ref{mnut4C}) satisfy the first law $%
dS=\beta dH$. In the flat space limit $I,S\rightarrow -4\pi N$, and go to
zero in the high temperature limit.

The specific heat $C=-\beta \partial _{\beta }S$ is 
\begin{equation}
C_{4C,NUT}=\frac{8\pi N^{2}(\ell ^{2}+12N^{2})}{\ell ^{2}}.  \label{Cnut4C}
\end{equation}%
The flat space and high temperature limits of the specific heat are $8\pi N$
and $0$, respectively.

The NUT solution is thus shown to satisfy the $\mathbf{N}$--bound for all $N$%
, as the entropy (\ref{Snut4C}) is always negative. We take this to mean
that the NUT solution is thermodynamically unstable everywhere. Note that
the specific heat is always positive, however.

\subsubsection{Bolt solution}

The Bolt solution has a two--dimensional fixed point set of $\partial _{T}$,
and thus $\rho _{+}=\rho _{b\pm }>N$. There are two conditions for a regular
Bolt solution:

\begin{description}
\item[(i)] $F(\rho) = 0 $

\item[(ii)] $F^{\prime }(\rho )=\pm \frac{q}{2N}$
\end{description}

\noindent where (ii) arises from the second equality in (\ref{beta4C}) and $
N>0$. The mass parameter in the Bolt case comes from (i), 
\begin{equation}
m_{4C,b}=-\frac{\left( \rho _{b}^{4}-(\ell ^{2}+6N^{2})\rho
_{b}^{2}-N^{2}(\ell ^{2}+3N^{2})\right) }{2\ell ^{2}\rho _{b}}.
\label{mbolt4C}
\end{equation}
The Bolt radii can be found from (ii); 
\begin{equation}
\rho _{b\pm }=\frac{q\ell ^{2}\pm \sqrt{q^{2}\ell ^{4}+48N^{2}\ell
^{2}+144N^{4}}}{12N}.  \label{rpm4C}
\end{equation}%
Note that the discriminant here is always greater than zero, so the only
restriction on the range of $N$ is $N>0$. Also, the flat space and high
temperature limits of $\rho _{b+}$ are infinite; the flat space limit of $%
\rho_{b-}$ is $-{\textstyle\frac{2N}{q}}$ and the high temperature limit is
zero.

The first equality in (\ref{beta4C}) gives the period for the Bolt as 
\begin{equation}
\beta _{4C,b}=2\pi \left| \frac{(\rho _{b}^{2}-N^{2})^{2}\ell ^{2}}{\rho
_{b}^{5}-2N^{2}\rho _{b}^{3}+N^{2}(9N^{2}+2\ell ^{2})\rho _{b}-m\ell
^{2}(\rho _{b}^{2}+N^{2})}\right| .  \label{betabolt4C}
\end{equation}%
The temperature of the NUT and Bolt solutions can be shown to be the same by
substituting in $m=m_{4C,b}$ and either of $\rho _{b}=\rho _{b\pm }$ into (%
\ref{betabolt4C}).

A plot of $m_{4C,b}$ for $\rho _{b}=\rho _{b\pm }$ is given in fig. \ref%
{plotmbolt4C}. 
\begin{figure}[tbp]
\centering       
\begin{minipage}[c]{.45\textwidth}
        \centering
        \includegraphics[width=\textwidth]{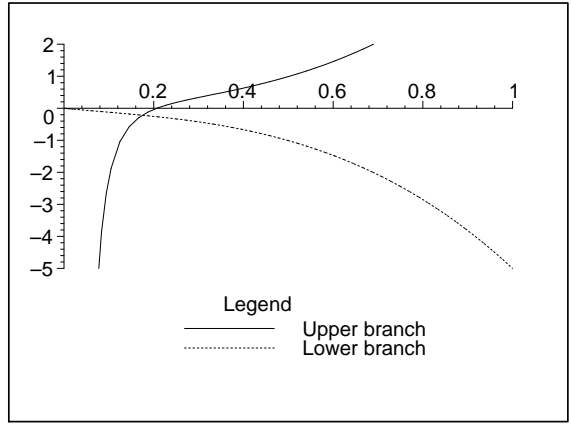}
    \end{minipage}
\caption{Plot of the upper ($\protect\rho _{b}=\protect\rho _{b+}$) and
lower ( $\protect\rho _{b}=\protect\rho _{b-}$) bolt masses $m_{b\pm }$ (for 
$q=1$) for 4 dimensions.}
\label{plotmbolt4C}
\end{figure}
Again from (\ref{mcons4dC}), $\mathfrak{M}=-m$, so from fig. \ref%
{plotmbolt4C} (and for $q=1$), we can see that the lower branch ($\rho
_{b}=\rho _{b-}$) mass will always be positive. This means that the lower
branch Bolt violates the maximal mass conjecture for all values of $N$. The
upper branch is negative (for $q=1$) for $N<0.2066200733$, and so violates
the conjecture for $N$ less than this. This trend holds for $q>1$, with the
cross-over point on the upper branch increasing with increasing $q$.

\begin{figure}[tbp]
\begin{center}
\begin{minipage}[c]{.45\textwidth}
         \centering
         \includegraphics[width=\textwidth]{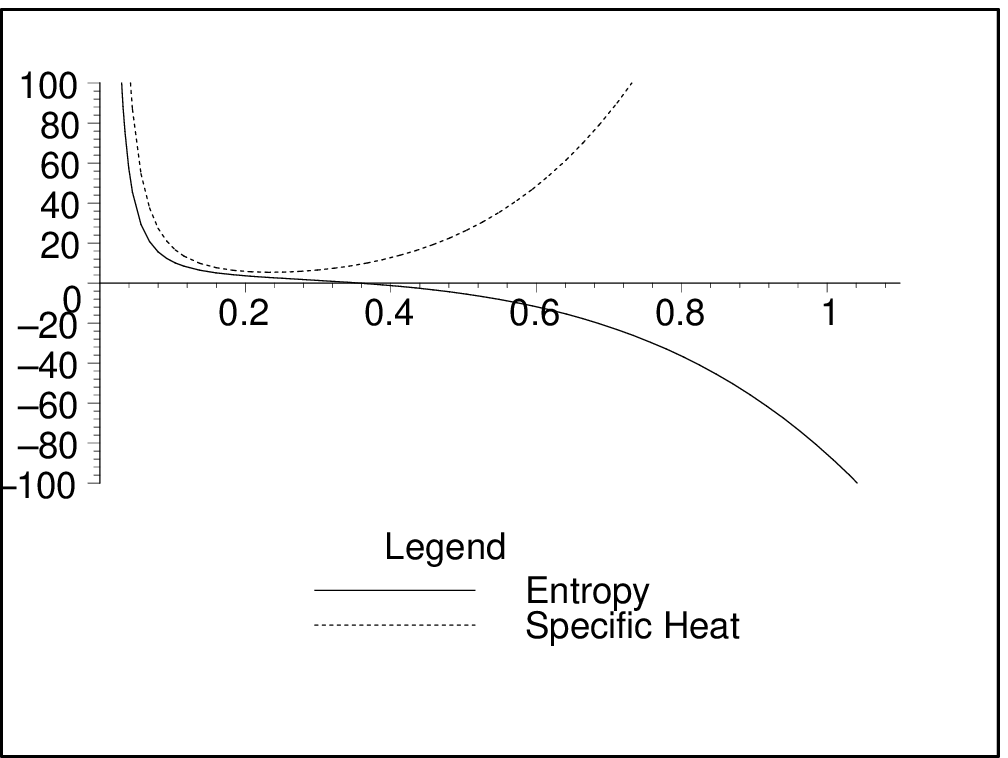}
\end{minipage}
\begin{minipage}[c]{.45\textwidth}
         \centering
         \includegraphics[width=\textwidth]{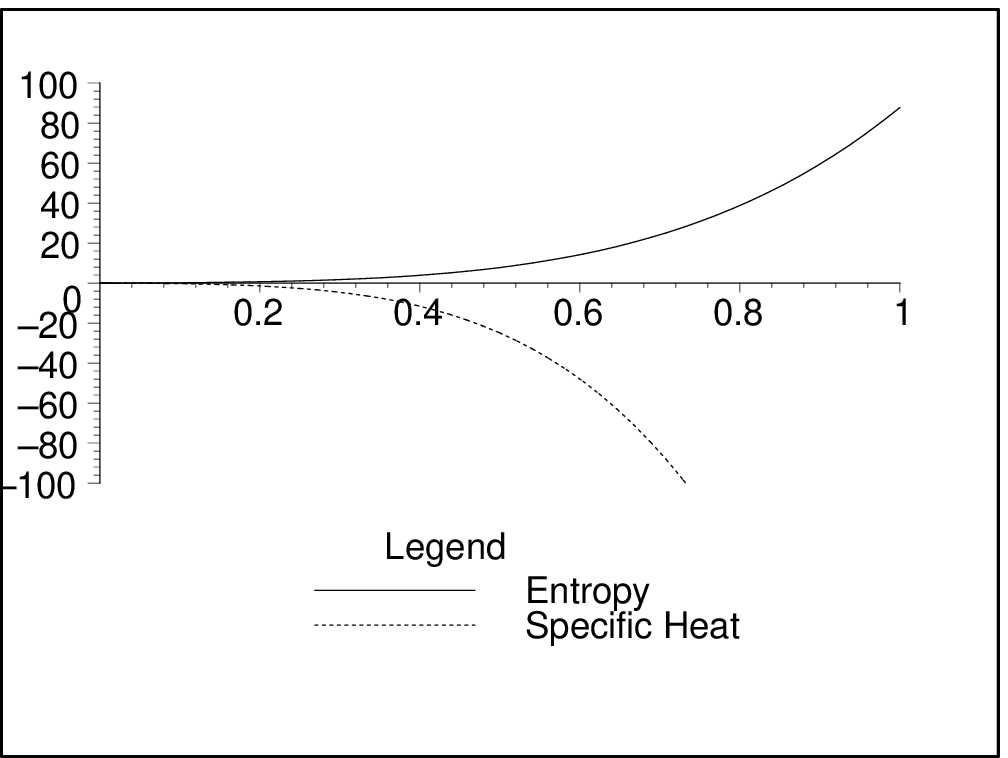}
\end{minipage}
\end{center}
\caption{Left/right: Plot of the upper/lower branch bolt entropy and
specific heat (for $q=1$) - Note that the entropy of pure de-Sitter space is 
$\protect\pi \ell ^{2}$.}
\label{plotSCpbolt4C}
\end{figure}

The action for the Bolt from (\ref{betabolt4C}) and (\ref{I4C}) is 
\begin{eqnarray}
I_{4C,b}(\rho _{b}=\rho _{b\pm }) &=&-\frac{\pi (\rho _{b}^{4}+\ell ^{2}\rho
_{b}^{2}+N^{2}(\ell ^{2}+3N^{2}))}{\rho _{b}}\left| \frac{\rho _{b}}{3\rho
_{b}^{2}-3N^{2}-\ell ^{2}}\right|  \label{Ibolt4C} \\
&=&-\frac{\pi }{216}\left[ \frac{(q^{2}\ell ^{2}+72N^{2})\ell ^{2}}{N^{2}}%
\pm \frac{(q^{2}\ell ^{4}+144N^{4}+48N^{2}\ell ^{2})^{3/2}}{qN^{2}\ell ^{2}}%
\right] .  \notag
\end{eqnarray}%
Similarly, the entropy is given by 
\begin{eqnarray}
S_{4C,b}(\rho _{b}=\rho _{b\pm }) &=&\frac{\pi (3\rho _{b}^{4}-(\ell
^{2}+12N^{2})\rho _{b}^{2}-N^{2}(\ell ^{2}+3N^{2}))}{\rho _{b}}\Bigg|\frac{%
\rho _{b}}{3\rho _{b}^{2}-3N^{2}-\ell ^{2}}\Bigg|  \notag \\
&=&\frac{\pi }{72}\left[ \frac{(q^{2}\ell ^{2}+24N^{2})\ell ^{2}}{N^{2}}%
\right.  \notag \\
&&\left. \pm \frac{(q\ell ^{2}-12N^{2})(q\ell ^{2}+12N^{2})\sqrt{q^{2}\ell
^{4}+144N^{4}+48N^{2}\ell ^{2}}}{\ell ^{2}qN^{2}}\right] .  \label{Sbolt4C}
\end{eqnarray}%
This can also be shown to satisfy the first law, by checking both $\rho
_{b}=\rho _{b\pm }$ cases separately. The specific heat is 
\begin{eqnarray}
C_{4C,b}(\rho _{b\pm }) & = & \frac{\pi }{36N^{2}}\Bigg[ q^{2}\ell^{4}
\label{Cbolt4C} \\
& & \pm \frac{ (144q^{2}\ell ^{4}N^{4}+41472N^{8}+10368N^{6}\ell
^{2}+24N^{2}\ell ^{6}q^{2}+q^{4}\ell ^{8})}{q\ell ^{2}\sqrt{q^{2}\ell
^{4}+144N^{4}+48N^{2}\ell ^{2}}}\Bigg].  \notag
\end{eqnarray}
The entropy and specific heat can be plotted; the upper and lower branch
plots are in fig. \ref{plotSCpbolt4C}. 
These show that the upper branch entropy is positive for $N<0.3562261982\ell 
$, and the specific heat is always positive. We take this to mean that the
upper branch is thermodynamically stable for $N$ less than this value.
However, the lower branch solution is thermodynamically unstable for all $N$%
, for while the entropy here is always positive, the specific heat is always
negative. This trend continues for $q>1$.

Figure \ref{plotSCpbolt4C} also shows that the Bolt solutions are
counter--examples of the $\mathbf{N}$--bound conjecture. The upper branch
entropy is greater than the pure de Sitter value for $N<0.2180098653\ell $,
and the lower branch entropy is greater than pure de Sitter for $%
N>0.3716679966\ell $, where the entropy of de Sitter is $\pi \ell ^{2}$.

\subsection{Six Dimensional Case}

The general metric in (5+1) dimensions, before specifying either a NUT or a
Bolt solution, and using $S^{2}\otimes S^{2}$ as a base space, takes the
form 
\begin{eqnarray}
ds_{6}^{2} &=&-F(\rho )\left[ dT+2N\cos (\theta _{1})d\phi _{1}+2N\cos
(\theta _{2})d\phi _{2}\right] ^{2}-\frac{d\rho ^{2}}{F(\rho )}  \notag \\
&&+(\rho ^{2}-N^{2})\left( d\theta _{1}^{2}+\sin ^{2}(\theta _{1})d\phi
_{1}^{2}+d\theta _{2}^{2}+\sin ^{2}(\theta _{2})d\phi _{2}^{2}\right)
\label{ds6C}
\end{eqnarray}%
where the metric function is given by 
\begin{equation}
F(\rho )=\frac{3\rho ^{6}-(\ell ^{2}+15N^{2})\rho ^{4}+3N^{2}(2\ell
^{2}+15N^{2})\rho ^{2}+3N^{4}(\ell ^{2}+5N^{2})+6m\rho \ell ^{2}}{3(\rho
^{2}-N^{2})^{2}\ell ^{2}}  \label{Fr6C}
\end{equation}%
with $N$ again the non--vanishing NUT charge, and $\Lambda ={\textstyle\frac{
10}{\ell ^{2}}}$. The periodicity condition to avoid conical singularities
in six dimensions is 
\begin{equation}
\beta _{C,6}=\frac{4\pi }{|F^{\prime }(\rho )|}=\frac{12\pi |N|}{q}.
\label{beta6C}
\end{equation}%
The geometric interpretation here is as murky as that in four--dimensions,
but we shall nevertheless proceed.

Working at future infinity, the action in six dimensions can again be found
from the counter--term method, giving 
\begin{equation}
I_{C,6}=-\frac{2\pi \beta \left( 3\rho _{+}^{5}+15N^{4}\rho _{+}+3m\ell
^{2}-10N^{2}\rho _{+}^{3}\right) }{3\ell ^{2}}  \label{I6dC}
\end{equation}%
where $\rho _{+}$ is again the largest positive root of $F(\rho )$. The
conserved mass is also found through the counter--term method, giving 
\begin{equation}
\mathfrak{M}_{C,6}=-8\pi m_{C,6d}-\frac{\pi (63\ell ^{4}N^{2}+2205N^{4}\ell
^{2}+10773N^{6}-\ell ^{6})}{54\rho \ell ^{2}}+\mathcal{O}\left( \frac{1}{%
\rho ^{2}}\right)  \label{Mcons6dC}
\end{equation}%
and the total entropy is found using these and the Gibbs--Duhem relation 
\begin{equation}
S_{C,6}=\frac{2\pi \beta (3\rho _{+}^{5}-10N^{2}\rho _{+}^{3}+15N^{4}\rho
_{+}-9m\ell ^{2})}{3\ell ^{2}}.  \label{S6dC}
\end{equation}%
The above equations are generic to six--dimensions, and can be further
analyzed by specifying to either the six dimensional Taub--NUT--C solution,
where $\rho_{+}=N, F(\rho =N)=0$ and the fixed point set of $\partial_{T}$
is 2-dimensional, or the Taub--Bolt--C solutions, where $\rho_{+}=\rho_{b\pm
}>N$ and the fixed point set of $\partial_{T}$ is 4--dimensional.

\subsubsection{NUT Solution}

The NUT mass is given by 
\begin{equation}
m_{C,n6} = -\frac{4N^3 (\ell^2 + 6N^2)}{\ell^2}  \label{mass6C}
\end{equation}
and is obviously always negative. This means, from (\ref{Mcons6dC}), that
the conserved mass at future infinity in six dimensions is always positive.
In the flat space limit, $m_{C,n6} \rightarrow -{\textstyle\frac{4}{3}}N^3$,
and in the high--temperature limit, $m_{C,n6} \rightarrow 0$.

The period here is $\beta _{6d}=12\pi N$, and so 
\begin{eqnarray}
I_{C,6d~NUT} &=&\frac{32\pi ^{2}N^{4}(\ell ^{2}+4N^{2})}{\ell ^{2}}
\label{INUT6dC} \\
S_{C,6d~NUT} &=&\frac{32\pi ^{2}N^{4}(3\ell ^{2}+20N^{2})}{\ell ^{2}}
\label{SNUT6dC}
\end{eqnarray}%
(\ref{SNUT6dC}), (\ref{Mcons6dC}) and (\ref{mass6C}) can be shown to satisfy
the first law $dS=\beta d\mathfrak{M}$. Both the action and the entropy are
zero in the high temperature limit. The flat space limits are $
I_{C,6d~NUT}\rightarrow 32\pi ^{2}N^{4}$, $S_{C6d~NUT}\rightarrow 96\pi
^{2}N^{4}$.

The specific heat in six dimensions is 
\begin{equation}
C_{C,6d ~NUT} = -\frac{384 \pi^2 N^4 (\ell^2 + 10N^2)}{\ell^2}
\label{CNUT6dC}
\end{equation}
which $\rightarrow -384 \pi^2 N^4$ and $\rightarrow 0$ in the flat space and
high temperature limits, respectively.

\begin{figure}[tbp]
\centering       
\begin{minipage}[c]{.45\textwidth}
         \centering
         \includegraphics[width=\textwidth]{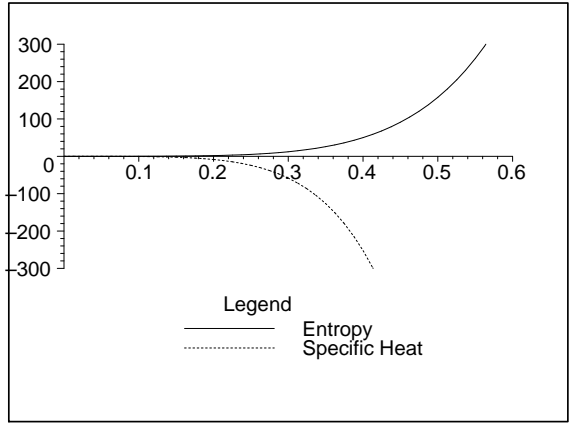}
         \end{minipage}
\caption{Plot of the NUT entropy and specific heat vs. $N$ for (5+1)
dimensions.}
\label{PlotSCNUT6d}
\end{figure}

Note that as shown in fig. \ref{PlotSCNUT6d}, the entropy and specific heat
behave opposite to the behaviour noted in four dimensions, with the entropy
always positive, and the specific heat negative. This implies that the six
dimensional NUT solution is also thermodynamically unstable. 

\subsubsection{Bolt solution}

Here the fixed point set of $\partial_{T}$ is four--dimensional. The
conditions for a regular Bolt solution are now

\begin{description}
\item[(i)] $F(\rho) = 0 $

\item[(ii)] $F^{\prime}(\rho) = \pm \frac{q}{3N}$
\end{description}

\noindent (with (ii) from the second equality in (\ref{beta6C})). From (i),
we can find the Bolt mass 
\begin{equation}
m_{C,b6}=-\frac{3\rho _{b}^{6}-(\ell ^{2}+15N^{2})\rho _{b}^{4}+N^{2}(6\ell
^{2}+45N^{2})\rho _{b}^{2}+3N^{4}(\ell ^{2}+5N^{2})}{6\ell ^{2}\rho _{b}}
\label{massbolt6C}
\end{equation}%
and from (ii) we can find $\rho _{b\pm }$ 
\begin{equation}
\rho _{b\pm }=\frac{q\ell ^{2}\pm \sqrt{q^{2}\ell ^{4}+900N^{4}+180N^{2}\ell
^{2}}}{30N}.  \label{rhopmbolt}
\end{equation}%
Note that, as in four dimensions, the discriminant of the square--root in $
\rho _{b\pm }$ is always positive, and so the only limit on $N$ is $N>0$.
The flat space and high temperature limits of $\rho _{b+}$ are both $\rho
_{b+}\rightarrow \infty $, and the flat space and high temperature limits of 
$\rho _{b-}$ are $\rho _{b-}\rightarrow -{\textstyle\frac{3N}{q}}$, $\rho
_{b-}\rightarrow 0$, respectively.

The period of the Bolt is found from the first equality in (\ref{beta6C}) 
\begin{eqnarray}
\beta _{C,Bolt6d} & = & 6\pi \Bigg| (\rho _{b}^{2}-N^{2})^{3}\ell^{2} \Bigg( %
3\rho_{b}^{7}-9\rho_{b}^{5}N^{2}-N^{2}(4\ell^{2}+15N^{2})\rho_{b}^{3}  \notag
\\
& & -9m\ell^{2}\rho _{b}^{2}-N^{4}(12\ell ^{2}+75N^{2})\rho _{b}-3m\ell
^{2}N^{2} \Bigg)^{-1} \Bigg| .  \label{betabolt6C}
\end{eqnarray}
This temperature is again the same as in the NUT case, as can be seen by
substituting in $m=m_{C,b6}$ and either $\rho _{b}=\rho _{b\pm }$.

\begin{figure}[tbp]
\centering       
\begin{minipage}[c]{.45\textwidth}
         \centering
         \includegraphics[width=\textwidth]{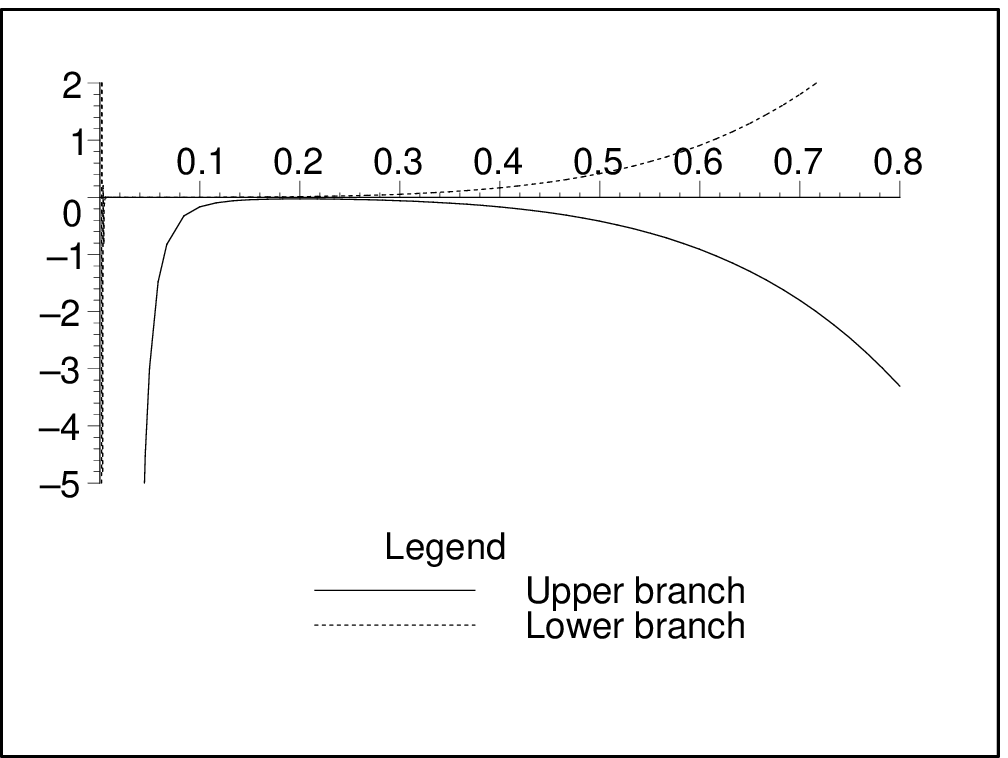}
    \end{minipage}
\caption{Plot of the upper ($\protect\rho _{b}=\protect\rho _{b+}$) and
lower ($\protect\rho _{b}=\protect\rho _{b-}$) bolt masses $m_{b\pm }$ ($q=1$%
) for six dimensions (Taub-Bolt-C).}
\label{Plotmbpm6d}
\end{figure}

The upper and lower branch masses for six dimensions can be plotted vs. $N$,
as is done in fig. \ref{Plotmbpm6d}. 
Note the upper branch is always negative, the lower is always positive.
Since $\mathfrak{M}_{C,6} = -8\pi m_{C,6d}$ this means the upper (lower)
branch conserved mass is always positive (negative). This behaviour is
different from that in four dimensions, where the upper branch varied from
positive to negative.

The action can be found from the method of counter--terms to be 
\begin{eqnarray}
I_{C,Bolt6d} &=&-\frac{4\pi ^{2}(3\rho _{b}^{6}+(\ell ^{2}-5N^{2})\rho
_{b}^{4}-N^{2}(6\ell ^{2}+15N^{2})\rho _{b}^{2}-3N^{4}(\ell ^{2}+5N^{2}))}{%
3|5N^{2}-5\rho _{b}^{2}+\ell ^{2}|}  \notag \\
&=&\frac{2\pi ^{2}}{253125}\left[ -\frac{\ell
^{4}(20250N^{4}+750N^{4}q^{2}+300q^{2}\ell ^{2}N^{2}+q^{4}\ell ^{4})}{N^{4}}%
\right.  \notag \\
&&\left. \pm \frac{(-q^{2}\ell ^{4}+600N^{4}-30N^{2}\ell ^{2})(q^{2}\ell
^{4}+900N^{4}+180N^{2}\ell ^{2})^{3/2}}{\ell ^{2}qN^{4}}\right]
\label{Ibolt6C}
\end{eqnarray}%
and, using the Gibbs--Duhem relation, 
\begin{eqnarray}
S_{C,Bolt6d} &=&\frac{4\pi ^{2}(15\rho _{b}^{6}-(3\ell ^{2}+65N^{2})\rho
_{b}^{4}+3N^{2}(6\ell ^{2}+55N^{2})\rho _{b}^{2}+9N^{4}(\ell ^{2}+5N^{2}))}{%
3|5N^{2}-5\rho _{b}^{2}+\ell ^{2}|}  \notag \\
&=&\frac{2\pi ^{2}}{50625}\Bigg[\frac{(q^{4}\ell ^{4}+180N^{2}q^{2}\ell
^{2}+150N^{4}q^{2}+4050N^{4})\ell ^{4}}{N^{4}}  \notag \\
&&\pm \frac{\sqrt{q^{2}\ell ^{4}+900N^{4}+180N^{2}\ell ^{2}}}{N^{4}q\ell ^{2}%
}(q^{4}\ell ^{8}+90N^{2}q^{2}\ell ^{6}-300N^{4}q^{2}\ell ^{4}  \notag \\
&&+27000N^{6}\ell ^{2}+540000N^{8})\Bigg].  \label{Sbolt6C}
\end{eqnarray}%
This entropy does satisfy the first law, though again both the upper and
lower branches must be checked separately. The specific heat is 
\begin{eqnarray}
C_{C,Bolt6d} &=&\frac{8\pi ^{2}}{50625}\left[ \frac{\ell
^{6}q^{2}(90N^{2}+q^{2}\ell ^{2})}{N^{4}}\right.  \notag \\
&&\left. \pm \frac{(q\ell ^{2}+30N^{2})(q\ell ^{2}-30N^{2})}{N^{4}q\ell ^{2}%
\sqrt{q^{2}\ell ^{4}+900N^{4}+180N^{2}\ell ^{2}}}\Big(q^{4}\ell
^{8}+180N^{2}q^{2}\ell ^{6}+1350N^{4}q^{2}\ell ^{4}\right.  \notag \\
&&\left. +4050N^{4}\ell ^{4}+162000N^{6}\ell ^{2}+810000N^{8}\Big)\right] .
\label{CCBolt6}
\end{eqnarray}

\begin{figure}[tbp]
\begin{center}
\begin{minipage}[c]{.45\textwidth}
         \centering
         \includegraphics[width=\textwidth]{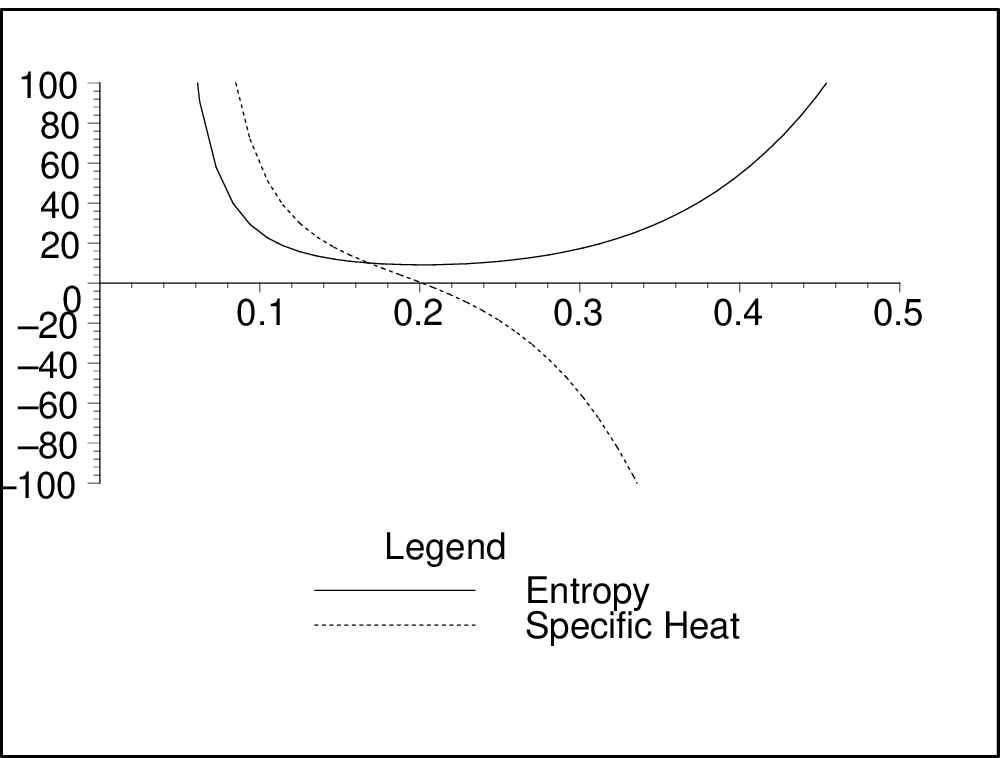}
    \end{minipage}
\begin{minipage}[c]{.45\textwidth}
         \centering
         \includegraphics[width=\textwidth]{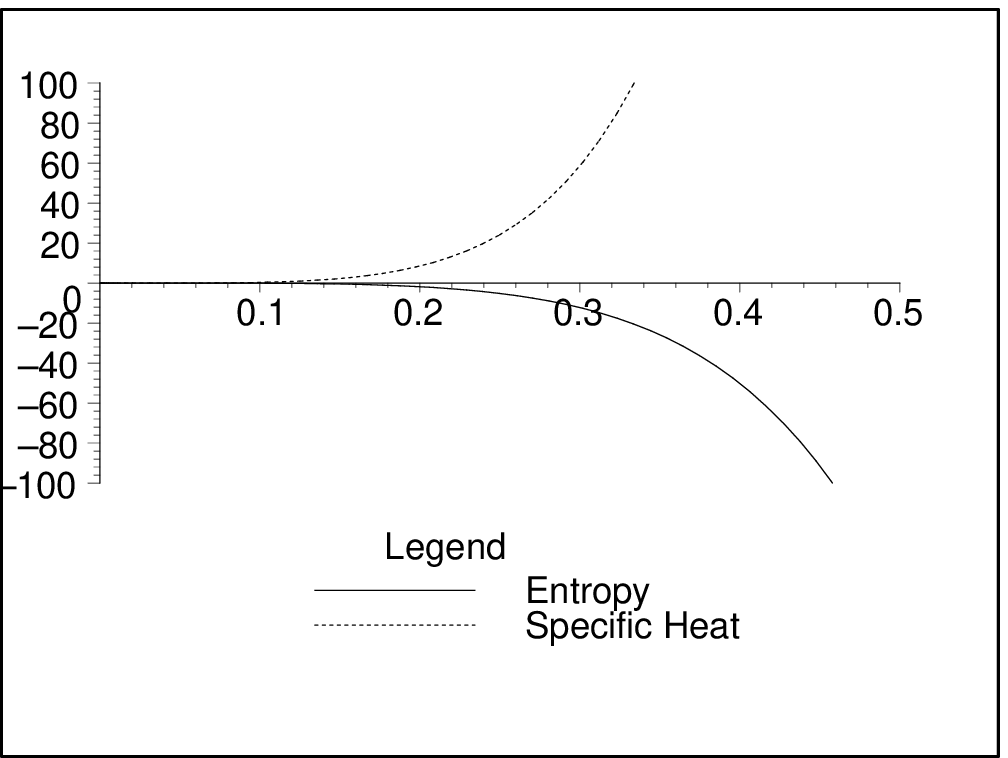}
    \end{minipage}
\end{center}
\caption{Left/right: Plot of the upper/lower branch bolt entropy and
specific heat (for $q=1$) for six dimensions. }
\label{PlotSCBoltp6d}
\end{figure}

Figure \ref{PlotSCBoltp6d} plots the entropy and specific heat for the upper
and lower branch solutions for $q=1$. The upper branch entropy is always
positive, and the specific heat is positive for $N<0.2014312523\ell$; thus
the upper branch solution is thermodynamically stable for $N$ less than
this. The lower branch entropy is always negative and specific heat always
positive, so the lower branch is always thermodynamically unstable. Note the
trend continues for $q>1$. 

\section{General $(d+1)$ Dimensional Results}

\label{sec:generalcalc}

The results in sections \ref{sec:examples} and \ref{sec:examplesC} can be
generalized to any even dimension greater than 4, where $d+1 = 2k+2$, ($%
k=1,2, \ldots$). For simplicity, we consider a base space of a product of
2-spheres $\otimes_{i=1}^k S^2$; note that this gives the same results as
would be obtained by using the more general case (\ref{TNDSgeneral}), and
that all of the general discussions regarding the structure, etc., of
Taub--NUT--dS spacetimes from section \ref{sec:cosmo-spaceimes} apply here.

We find that the behaviour of both the R--approach and C--approach
quantities are qualitatively the same in $4k$--dimensions ($k=1,2,\ldots $),
a behaviour that is distinct from the common behaviour in $4k+2$-dimensions.
This means that dimensions $8,12,16,\ldots $ have the same behaviour as four
dimensions, and $10,14,18,\ldots $ the same behaviour as six dimensions. We
have explicitly checked that this holds up to 20 dimensions.

\subsection{R--approach}

The general form of the metric for the R--approach using a product of
two--spheres is given by 
\begin{equation}
ds_R^2 = V(\tau) \left( dt + 2n \sum_{i=1}^k \cos (\theta_i) d\phi_i
\right)^2 - \frac{d\tau^2}{V(\tau)} + (\tau^2 + n^2) \sum_{i=1}^k \left(
d\theta_i^2 + \sin^2 (\theta_i) d\phi_i^2 \right)  \label{dsgeneralR}
\end{equation}
with $V(\tau)$ given by (\ref{FtBoltgen}). In the following we will denote
the largest root of $V(\tau)$ by $\tau_c$, where the subspace $\tau = \tau_c$
is the fixed point set of $\partial_t$.

The Killing vectors $\partial _{\phi _{i}}$ will give conical singularities
in the $(t,\tau )$ Euclidean section for any constant $(\phi _{1},\ldots
,\phi _{i})$--slice near the horizon $\tau =\tau _{c}$ unless we fix the
period to be 
\begin{equation}
\beta _{R}=\frac{4\pi }{|V^{\prime }(\tau )|}  \label{betageneralR}
\end{equation}%
This must match the periodicity requirement induced by requiring the Misner
string singularities vanish, giving the relation 
\begin{equation}
\frac{1}{|V^{\prime }(\tau )|}=\frac{(d+1)|n|}{2q}  \label{consistencygenR}
\end{equation}%
which will have two solutions for $\tau _{c}=\tau ^{\pm }(n)$. The fixed
point set of $\partial _{t}$ is $(d-1)$--dimensional for each of these, and
so both solutions are Bolt--solutions. We refer to these solutions as $R^{+}$
and $R^{-}$ respectively.

The metric determinant and Ricci scalar in $(d+1)$ dimensions can be
calculated from (\ref{dsgeneralR}) to be 
\begin{eqnarray}
g_{R} &=&-(\tau ^{2}+n^{2})^{(2k)}\prod_{i=1}^{k}\sin ^{2}(\theta _{i}) \\
R_{R} &=&\frac{d(d+1)}{\ell ^{2}}.
\end{eqnarray}%
With these, it is possible to calculate the general form of the Bulk action
for arbitrary $d$, (where $\prod \sin ^{2}(\theta _{i})$ will contribute to
the volume term). We work in the $\tau >\tau _{c}$ region near future
infinity, and can use the binomial expansion on the integrand to allow
integration term by term from $\tau =\tau _{c}\rightarrow \infty $. 
\begin{eqnarray}
I_{R,B} &=&\frac{\beta d(4\pi )^{k}}{8\pi \ell ^{2}}\int d\tau ~\left( \tau
^{2}+n^{2}\right) ^{k}  \notag \\
&=&-\frac{\beta d(4\pi )^{k}}{8\pi \ell ^{2}}\sum_{i=0}^{k}\binom{k}{i}n^{2i}%
\frac{\tau _{c}^{2k-2i+1}}{2k-2i+1}.  \label{IbulkgeneralR}
\end{eqnarray}%
Since we know (see below) that the counter--term action will cancel
infinities in this and the boundary action, we only need to include finite
contributions here. The boundary contributions at future infinity can be
found using the boundary metric $\gamma _{\mu \nu }=g_{\mu \nu }+n_{\mu
}n_{\nu }$, where $n_{\mu }=\left[ 0,{\textstyle\frac{1}{\sqrt{-g^{\tau \tau
}}}},0,\ldots \right] $ is the unit normal to a surface of fixed $\tau $.
The boundary metric determinant and boundary Ricci scalar are, for general $%
d $ 
\begin{eqnarray}
\gamma _{R} &=&V(\tau )(\tau ^{2}+n^{2})^{2k}\prod_{i=1}^{k}\sin ^{2}(\theta
_{i})  \label{gboundarygeneralR} \\
R_{R}(\gamma ) &=&(d-1)\left[ \frac{1}{(\tau ^{2}+n^{2})}-\frac{V(\tau )n^{2}%
}{(\tau ^{2}+n^{2})^{2}}\right] .  \label{RicciboundarygeneralR}
\end{eqnarray}%
The trace of the extrinsic curvature for general $d$ can also be found from
the metric (\ref{dsgeneralR}) to be 
\begin{equation}
\Theta _{R}=-\left[ \frac{V^{\prime }(\tau )}{2\sqrt{V(\tau )}}+\frac{%
(d-1)\tau \sqrt{V(\tau )}}{(\tau ^{2}+n^{2})}\right] .  \label{ThetageneralR}
\end{equation}%
Equations (\ref{gboundarygeneralR}) and (\ref{ThetageneralR}) can be
expanded in (\ref{actbound}) for large $\tau $, and the finite terms can be
extracted. The finite boundary action is then 
\begin{equation}
I_{R,\partial B\text{finite}}=-\frac{\beta (4\pi )^{k}d}{8\pi }m.
\label{IboundarygeneralR}
\end{equation}

The counter--term action (\ref{counter}), (\ref{count345}) is now used to
cancel the divergent terms in the bulk and boundary actions, but will also
contribute finite terms to the total action. It turns out, however, that the
finite terms from the counter--term action come only from the first term in (%
\ref{count345}). This can be understood as follows \cite{CFM}. From (\ref%
{FtBoltgen}), we can expand 
\begin{equation}
V(\tau )=-\sum_{j}\tau ^{2p_{j}}-\frac{2m}{\tau ^{2k+1}}+\text{terms that
vanish as }\tau \rightarrow \infty
\end{equation}%
where the $p_{j}>0$ are integers. Any term that depends on $m$ must also
depend on an odd power of $\tau $. It therefore cannot cancel any
divergences, since from the above expansion, all divergences must be
cancelled by terms that are even in $\tau $ in the large $\tau $ limit.
Dimensionality requirements force all non--divergent counter--term
contributions depending on $m$ to be down by at least one power from $%
(2m)/(\tau ^{2k+1})$, and thus they must behave like $(2m)/(\tau ^{2k+2})$.
Since $\sqrt{\gamma }\sim \tau ^{2k+1}$ in the large $\tau $ limit, these
will all vanish. Likewise all other non--divergent counter--term
contributions must be down by two powers, and will also vanish upon
integration. We have checked these statements for $(d+1)=4\ldots 20$.
Therefore, the finite contribution from the counter--term action at future
infinity is 
\begin{equation}
I_{R,ct\text{finite}}=\frac{\beta (4\pi )^{k}(d-1)}{8\pi }m.
\label{IctgeneralR}
\end{equation}%
Adding the contributions (\ref{IbulkgeneralR}), (\ref{IboundarygeneralR})
and (\ref{IctgeneralR}) together, the general form of the action in the
R--approach is given by 
\begin{equation}
I_{R\text{ finite}}=-\frac{\beta (4\pi )^{k}}{8\pi }\left[ m+\frac{d}{\ell
^{2}}\sum_{i=0}^{k}\binom{k}{i}n^{2i}\frac{\tau _{c}^{2k-2i+1}}{2k-2i+1}%
\right] .  \label{ItotgenR}
\end{equation}

The conserved charge can also be found for general $d$. The only
non--vanishing conserved charge will be associated with $\xi =\partial _{t}$%
, and will give a conserved mass. Thus, using (\ref{Mcons}), we find 
\begin{equation}
\mathfrak{M}=\frac{1}{8\pi }\int d^{d-1}x~~\sqrt{\gamma }\left[ \Theta
_{ab}-\Theta \gamma _{ab}+\frac{(d-1)}{\ell }\gamma _{ab}+\ldots \right]
n^{a}\xi ^{b}.  \label{Mconsgen}
\end{equation}%
The extra terms from the variation of the counter--term action have been
previously calculated \cite{GM}, though using the arguments from above, we
have shown that only the first term ${\textstyle\frac{(d-1)}{\ell }}\gamma
_{ab}$ will give a finite contribution. The finite conserved mass for
general $d+1$ dimensions is thus given by 
\begin{equation}
\mathfrak{M}_{R}=-\frac{(4\pi )^{k}(d-1)}{8\pi }m  \label{MconsgenR}
\end{equation}%
$m$ can be solved for on a dimension by dimension basis in terms of $\tau ,n$%
, through the first condition $V(\tau )=0$.

The entropy in $d+1$ dimensions is given by 
\begin{equation}
S_{R}=\frac{(4\pi )^{k}\beta }{8\pi }\left\{ \frac{d}{\ell ^{2}}%
\sum_{i=0}^{k}\binom{k}{i}n^{2i}\frac{\tau _{C}^{2k-2i+1}}{2k-2i+1}%
-(2k-1)m\right\}  \label{SgenR}
\end{equation}%
where we use (\ref{MconsgenR}), (\ref{ItotgenR}) and the Gibbs-Duhem
relation.

Finally, note that the consistency condition (\ref{consistencygenR}) can be
rewritten 
\begin{equation}
|V^{\prime}(\tau_c)| = \frac{2q}{(d+1)|n|}  \label{consistencygenR2}
\end{equation}
and will yield four solutions for $\tau_c$ for a given dimension. Two of
these will be positive, and this implies two distinct spacetimes, each with
its own characteristic entropy and conserved mass for a given $n$.

\subsection{C--approach}

The metric in this approach is found from (\ref{dsgeneralR}) by a Wick
rotation of the time and the NUT charge ($t\rightarrow \text{i}T,
n\rightarrow \text{i}N$) to give 
\begin{equation}
ds_{C}^{2}=-F(\rho )\left( dT+2N\sum_{i=1}^{k}\cos (\theta _{i})d\phi
_{i}\right) ^{2}-\frac{d\rho ^{2}}{F(\rho )}+(\rho
^{2}-N^{2})\sum_{i=1}^{k}\left( d\theta _{i}^{2}+\sin ^{2}(\theta _{i})d\phi
_{i}^{2}\right)  \label{dsgeneralC}
\end{equation}%
with $F(\rho )$ found from (\ref{FtBoltgen}) 
\begin{equation}
F(\rho )=\frac{2m\rho }{(\rho ^{2}-N^{2})^{k}}-\frac{\rho }{(\rho
^{2}-N^{2})^{k}}\int_{\rho }ds\left[ \frac{(s^{2}-N^{2})^{k}}{s^{2}}-\frac{%
(2k+1)}{\ell ^{2}}\frac{(s^{2}-N^{2})^{k+1}}{s^{2}}\right]  \label{FtgenR}
\end{equation}%
This metric, up to a few signs, is the same as in the R--approach, and so
the same calculations can be used from the R--approach to find the
quantities needed to calculate the action, etc.. The general metric
determinant and Ricci scalar are thus given by 
\begin{eqnarray}
g_{C} &=&(\rho ^{2}-N^{2})\prod_{i=1}^{k}\sin (\theta _{i})  \label{ggenC} \\
R_{C} &=&\frac{d(d+1)}{\ell ^{2}}.  \label{RgenC}
\end{eqnarray}%
The finite bulk action contribution is found by inserting these into (\ref%
{actbulk}), with the binomial expansion, giving 
\begin{equation}
I_{C,B}=-\frac{(4\pi )^{k}\beta d}{8\pi \ell ^{2}}\sum_{i=0}^{k}\binom{k}{i}%
(-1)^{i}N^{2i}\frac{\rho _{+}^{2k-2i+1}}{2k-2i+1}  \label{IbulkC}
\end{equation}%
where $\rho _{+}$ is the largest positive root of $F(\rho )$, found using
the fixed point set of $\partial _{T}$, and $\beta $ here is the period of $%
T $, again found by ensuring regularity in the $(T,\rho )$ section via the
formula 
\begin{equation}
\beta _{C}=\frac{4\pi }{|F^{\prime }(\rho _{+})|}=\frac{2(d+1)\pi |N|}{q}
\label{betagenC}
\end{equation}%
It is important to note that $\rho _{+}\neq \tau _{C}$ due to the changes in
going from $V(\tau )$ to $F(\rho )$.

The boundary metric is again found through the formula $\gamma_{\mu \nu
}=g_{\mu \nu }+n_{\mu }n_{\nu }$, with $n_{\mu }$ as above, and we are again
working only at future infinity. This gives the boundary metric quantities 
\begin{eqnarray}
\gamma _{C} &=&-F(\rho )(\rho ^{2}-N^{2})^{2k}\prod_{i=1}^{k}\sin
^{2}(\theta _{i})  \label{gammadetC} \\
R_{C}(\gamma ) &=&(d-2)\left[ \frac{1}{(\rho ^{2}-N^{2})}+\frac{F(\rho
)N^{2} }{(\rho ^{2}-N^{2})^{2}}\right]  \label{RgammagenC}
\end{eqnarray}%
and from (\ref{dsgeneralC}) we can find 
\begin{equation}
\Theta _{C}=-\left[ \frac{F^{\prime }(\rho )}{2\sqrt{F(\rho )}}+\frac{%
(d-1)\rho \sqrt{F(\rho )}}{(\rho ^{2}-N^{2})}\right] .  \label{ThetagenC}
\end{equation}%
The general equations for the finite contributions from the boundary and
counter--term actions can be found to be the same here as in the R--approach
(\ref{IboundarygeneralR}), (\ref{IctgeneralR}), using exactly the same steps
--- including the fact that only the first term in (\ref{count345}) is
needed for (\ref{IctgeneralR}). Thus, the general form of the action for the
Taub--NUT--dS (C--approach) metric is given by 
\begin{equation}
I_{C\text{ finite}}=-\frac{(4\pi )^{k}\beta }{8\pi }\left[ m+\frac{d}{\ell
^{2}}\sum_{i=0}^{k}\binom{k}{i}(-1)^{i}N^{2i}\frac{\rho _{+}^{2k-2i+1}}{%
2k-2i+1}\right] .  \label{IgenC}
\end{equation}%
The conserved mass is also again found from (\ref{Mconsgen}), where again
only the first three terms given in (\ref{Mconsgen}) are needed. This gives
the finite conserved mass in general C--approach spacetimes as 
\begin{equation}
\mathfrak{M}_{C}=-\frac{(4\pi )^{k}2k}{8\pi }m  \label{MconsgenC}
\end{equation}%
(where recall that the total dimension $(d+1)=2k+2$).

The Gibbs--Duhem relation (\ref{GDoutfinal}) can now be used to find the
entropy 
\begin{equation}
S_{C}=\frac{(4\pi )^{k}\beta }{8\pi }\left[ \frac{d}{\ell ^{2}}\sum_{i=0}^{k}%
\binom{k}{i}(-1)^{i}N^{2i}\frac{\rho _{+}^{2k-2i+1}}{2k-2i+1}-(2k-1)m\right]
.  \label{SgenC}
\end{equation}%
The periodicity condition for ensuring regularity (\ref{betagenC}) and
removal of all Misner strings now yields 
\begin{equation}
|F^{\prime }(\rho _{+})|=\frac{2q}{(d+1)|N|}.  \label{consistencygenC2}
\end{equation}%
However, unlike in the R--approach, we now have two distinct classes of
solutions to (\ref{consistencygenC2}), depending on the co--dimensionality
of the fixed point set of $\partial _{T}$. The fixed point set of $\partial
_{T}$ can be $(d-1)$ dimensional or it can be of $(d-3)$--dimensionality.
The $(d-1)$ dimensional case yields a solution $\rho _{+}>N$ and is called a
Bolt solution --- we thus refer to this case as the Taub--Bolt--C solution;
and the $(d-3)$--dimensional case yields a solution $\rho _{+}=N$ and is
called a NUT solution --- we refer to this as the Taub--NUT--C solution. It
has recently been shown \cite{RCstelea} that this solution can be
analytically continued to a Taub--NUT--R solution.

The entropy defined by the Gibbs--Duhem relation (\ref{GDoutfinal}) would
appear to have the requisite properties: it is positive and monotonically
increasing with conserved mass for the Schwarzschild--de Sitter case, and
obeys the first law of thermodynamics for all cases we have considered so
far (indeed, since our definition is built on the path integral formalism,
it is hard to see how it could be otherwise). However the applicability of
the second law remains an outstanding problem: in what sense can we say that
the entropy always increases in any physical process in this context? Even
more intriguing is the relationship between this entropy and the underlying
degrees of freedom that it presumably counts.

\section{Conclusions}

Looking at the previous results from a wider perspective, it is clear that
it is possible to extend the concepts of conserved quantities, actions,
entropies and path--integral methods outside of cosmological horizons. We
have extended the use of the path--integral formalism to include quantum
correlations between timelike histories. By employing this formalism in the
semiclassical approximation we have been able to extend our notions of
conserved quantities (such as mass and angular momentum), actions and
entropies outside of cosmological horizons. Applying this formalism to
Schwarzschild--de Sitter spacetimes we find that the values of these
quantities are in accord with our physical expectations \cite{bala,cai,GM}.

However when we extend this formalism to NUT--charged spacetimes we find
that the situation is considerably modified. First, NUT--charged spacetimes
present us with two possible ways (the R--approach and the C--approach) in
which we can apply our formalism, depending on how the spacetime is
analytically continued. We could, of course, have simply used one or the
other of these methods, as it has been recently shown \cite{RCstelea} that
the two approaches are equivalent and interchangeable through analytic
continuation. However, we felt that it was more demonstrative to show both
methods, for those more familiar with the C--approach from AdS analysis.

We have also found that there exist broad ranges of parameter space for
which NUT--charged spacetimes violate both the maximal mass conjecture and
the $\mathbf{N}$--bound, in both four dimensions and in higher dimensions.

We find the thermodynamic behaviour for the $4k$--dimensional spacetimes to
be qualitatively similar, with the lower branch entropy always negative, and
the upper branch solutions always having a range of $n$ in which both the
entropy and the specific heat are positive. Likewise the $(4k+2)$%
--dimensional spacetimes have qualitatively similar thermodynamics, behaving
as illustrated in figures (\ref{mass6pos}) and (\ref{ent6pos}).

The entropy--area relation $S=A/4$ is satisfied for any black hole in a $%
(d+1)$ dimensional aF, where $A$ is the area of a $d-1$ dimensional fixed
point set of isometry group. However, this relationship does not hold for
other kinds of spacetimes in which the isometry group has fixed points on
surfaces of even co-dimension \cite{TNUT}. The best examples of these
spacetimes are asymptotically locally flat and asymptotically locally AdS
spacetimes with NUT charge. In these cases when the isometry group has a
two--dimensional fixed set (Bolt), the entropy of the spacetime is not given
by the area--entropy relation, as a consequence of the first law of
thermodynamics \cite{as}.

In asymptotically dS spacetimes, the Gibbs--Duhem entropy (\ref{GDoutfinal})
is proportional to the area of the horizon and respects the $\mathbf{N}$%
--bound (this has also been shown for the Schwarzschild--dS spacetime \cite%
{GM}). However, for asymptotically locally dS spacetimes with NUT charge,
the entropy is no longer proportional to the area. Consequently the entropy
need not respect the $\mathbf{N}$--bound, and we find that there are a wide
range of situations where it does not.

It would be interesting to understand the range of applicability of the $%
\mathbf{N}$--bound is asymptotically dS spacetimes. Does it hold only for
spacetimes in which entropy is proportional to area? If not, what are the
minimal requirements for the $\mathbf{N}$--bound to hold? Can one make sense
of a theory of quantum gravity in which the $\mathbf{N}$--bound does not
hold? These questions remain interesting avenues for further study.

\section{Addendum}

We summarize here some developments that have taken place since the work
described in this paper was completed.One comment that has been raised
regarding the counterexamples we present has to do with the presence of
closed timelike curves (CTC's). \ The Chronological Protection Conjecture %
\cite{HawkingCPC} (CPC) suggests that spacetimes with CTC's develop
singularities upon perturbation of the stress tensor, and it has been
claimed \cite{BalaCPC} that the counter-examples we present are not true
counter-examples but are at best marginal. Following up on this, Anderson %
\cite{Anderson} obtained an counterexample to the maximal mass conjecture
using NUT-charged spacetimes that do not have CTC's and whose overall global
structure is the same as that of pure de Sitter space. This example can be
found from our four-dimensional Taub-NUT-dS metric (9.1) by analytic
continuation; the parameters of the metric as such that it excludes horizons
and so also excludes CTC's. \ 

We comment that our calculation of the conserved \ $(d+1)$-dimensional mass
from eq. (3.13) does not depend upon the existence of horizons (or CTCs).
Since the mass and NUT charge are a-priori independent, it is
straightforward to choose values of these quantities (in units of, say, the
cosmological constant) that preserve the global structure of pure de Sitter
space and violate the maximal mass conjecture. For example, one can choose
the values employed by Anderson \cite{Anderson}, recovering the
counterexample he presents. \ We also note that the applicability of the CPC
here requires some care; for example the maximal extension of Kerr-dS
spacetimes contains CTCs, yet their exclusion from some form of the maximal
mass conjecture would seem unduly restrictive. \ Of course a proof of the
maximal mass conjecture under reasonable physical restrictions remains an
interesting area of research.

Shortly after the completion of this review paper, we were made aware of
several papers of interest. Cai and Ohta \cite{CaiOhta} have discussed an
adaptation of the counterterm method for non-Anti de Sitter spacetimes, and
Cai, Myung and Ohta \cite{CaiMyungOhta} discuss the N-bound relative to
black holes. We also note that Nojiri and Odintsov \cite{NojOdint1}
constructed surface terms from the de Sitter bulk and obtained the 4
dimensional conformal anomaly; they also extended this asymptotically dS
bulk spacetimes and discussed the relationship with holographic RG flow \cite%
{NojOdint2}.

\section{Acknowledgements}

This work was supported by the Natural Sciences and Engineering Research
Council of Canada.

\end{document}